\documentclass[12pt]{iopart}

\expandafter\let\csname equation*\endcsname\relax
\expandafter\let\csname endequation*\endcsname\relax
\usepackage{amsmath}
\usepackage{amssymb}
\usepackage{graphicx}
\usepackage{booktabs}
\usepackage{multirow}
\usepackage{xcolor}
\usepackage{url}
\usepackage{algorithm}
\usepackage{algpseudocode}

\usepackage{cite}
\usepackage{chemformula} % Formula subscripts using \ch{}

\begin{document}

\title[]{Beyond Adam: Disentangling Optimizer Effects in the Fine-Tuning of Atomistic Foundation Models}

\author{Xiaoqing Liu}

\address{Department of Mechanical Engineering, National University of Singapore, 117575 Singapore.}

\author{Yangshuai Wang}

\address{Department of Mathematics, National University of Singapore, 10 Lower Kent Ridge Road, Singapore.}

\author{Teng Zhao}

\address{Shanghai Jiao Tong University–Chongqing Institute of Artificial Intelligence, Chongqing 401329, China.}

\ead{yswang@nus.edu.sg}
\ead{zhaoteng\_sjtu@sjtu.edu.cn}
\vspace{10pt}
% \begin{indented}
% \item[]August 2017
% \end{indented}

\begin{abstract}
Atomistic foundation models constitute a paradigm shift in computational materials science by providing universal machine-learned interatomic potentials with broad transferability across chemical spaces. Although fine-tuning is essential for adapting these pretrained models to specific target systems, the influence of the optimization algorithm on this process remains insufficiently characterized. In this work, we perform a rigorous benchmark of seven first-order optimizers, including Adam, AdamW, RAdam, SGD, LAMB, Ranger, and ScheduleFree, for the fine-tuning of foundation models across molecular, crystalline, and liquid regimes. We evaluate these algorithms based on energy and force accuracy for both in-distribution and out-of-distribution configurations, as well as their impact on downstream physical properties such as elastic moduli, phonon spectra, and interfacial dynamics. We interpret these empirical results through a preconditioning framework that views each optimizer as a data-dependent linear transformation of the gradient. This analysis clarifies how different update rules impose specific spectral filters on the effective loss Hessian. Across all regimes, AdamW and ScheduleFree achieve superior curvature conditioning and force accuracy, whereas stochastic gradient descent exhibits slow convergence and instability. Furthermore, we demonstrate that a brief second-order refinement stage reduces residual anisotropy in the loss landscape and enhances the fidelity of physical observables without increasing inference costs. These findings provide conceptual insight and practical guidance for selecting and designing optimizers to ensure the stable and efficient fine-tuning of universal interatomic potentials.
\end{abstract}
%
% Uncomment for keywords
%\vspace{2pc}
%\noindent{\it Keywords}: XXXXXX, YYYYYYYY, ZZZZZZZZZ
%
% Uncomment for Submitted to journal title message
%\submitto{\JPA}
%
% Uncomment if a separate title page is required
%\maketitle
% 
% For two-column output uncomment the next line and choose [10pt] rather than [12pt] in the \documentclass declaration
%\ioptwocol
%

\section{Introduction}
\label{sec:intro}

Machine-learned interatomic potentials (MLIPs) have emerged as an indispensable tool for atomistic modeling, enabling near density functional theory (DFT) accuracy at a fraction of computational cost~\cite{behler2007generalized, bartok2010gaussian, thompson2015spectral, shapeev2016moment, schutt2017schnet, smith2017ani, wang2018deepmd, DrautzACE, batatia2022mace, batzner20223, musaelian2023learning, xie2023ultra, bochkarev2024graph, cheng2024cartesian}. Various frameworks have been developed, including neural network potentials~\cite{behler2007generalized, smith2017ani, wang2018deepmd}, kernel-based approaches~\cite{bartok2010gaussian, DrautzACE}, and equivariant graph neural networks~\cite{batatia2022mace, batzner20223}. Comprehensive reviews of MLIPs can be found in~\cite{botu2017machine, unke2021machine, musil2021physics, poltavsky2021machine, jacobs2025practical}.

Traditional MLIPs are typically trained on narrowly defined chemical or structural domains. They achieve high accuracy in-distribution configurations, but their performance often degrades when extrapolated to unseen compositions, phases, or thermodynamic states. To address this limitation, recent work has introduced general purpose foundation or universal MLIPs (U-MLIPs)~\cite{batatia2023foundation, deng2023chgnet, merchant2023scaling, zhang2024dpa, choudhary2023unified, chen2022universal}, which are pre trained on large and diverse corpora of atomic environments~\cite{chanussot2021open, bowman2022md17, barroso2024open}. Models such as MACE-MP-0~\cite{batatia2023foundation}, CHGNet~\cite{deng2023chgnet}, MatterSim~\cite{yang2024mattersim}, EquiformerV2~\cite{barroso2024open}, and DPA~\cite{zhang2022dpa, zhang2024dpa} capture broad chemical interactions, and can then be adapted to new systems through fine-tuning with reduced data and computational cost.

Fine-tuning is therefore a critical step for turning universal representations into task-ready interatomic models. By refining a pre-trained backbone on system-specific data, fine-tuning narrows the gap between broad generality and targeted accuracy~\cite{focassio2024performance, deng2024overcoming, yu2024systematic, pyzer2025foundation, shuang2025universal, du2025universal, lee2025accelerating, niblett2024transferability, casillas2024evaluating, radova2025fine, liu2025fine, liu2025study}. Unlike large-scale pretraining, which is primarily driven by data diversity and computational throughput, the effectiveness of fine-tuning is strongly governed by the choice of optimizer, because the available data are more limited, the loss landscape is more anisotropic, and training budgets are tighter~\cite{wu2017towards}. As a result, curvature conditioning, step-size control, and implicit regularization induced by distinct optimization dynamics become key determinants of stability and transfer performance. Recent scaling studies further show that attention-based neural network interatomic potentials can increase expressivity while reducing inference time and memory, which, in turn, amplifies the need for optimization strategies that maintain stable and efficient convergence during fine-tuning~\cite{qu2024importance}. 

Despite this central role, the optimizer choice is often treated as a fixed default in MLIP workflows. The overwhelming majority of MLIP training andfine-tuning studies rely on Adam~\cite{kingma2014adam}, and only rarely motivate this choice beyond established convention~\cite{qi2024robust, anstine2023machine}. A few recent works introduce or compare alternative optimizers for neural network potentials, for example CoRe in lifelong machine learning potentials~\cite{eckhoff2023lifelong}, Kalman filter based schemes, and RLEKF for Deep Potentials~\cite{hu2023rlekf}, but these studies remain method specific rather than providing a broad benchmark across architectures and materials domains. In contrast, other areas of scientific and mainstream machine learning have begun to compare optimization algorithms in controlled settings, including numerical weather prediction, quantum machine learning, computer vision, and large language models, as well as general deep learning benchmarks that highlight the methodological subtleties of optimizer evaluation~\cite{llugsi2021comparison, vu2025benchmarking, hassan2023effect, semenov2025benchmarking, choi2019empirical}.

For atomistic foundation models, the optimizer controls how parameters move through the loss landscape and how the pre-trained energy manifold deforms to accommodate new environments~\cite{dozat2016incorporating, kingma2014adam}. Its dynamics govern the stability of convergence, the trade off between energy and force errors, and the smoothness of the learned potential energy surface. Different algorithms implement different mechanisms for the estimation of the curvature, gradient normalization, and adaptive rescaling, which can be viewed as different preconditioning operators that impose characteristic spectral filters on the Hessian of the loss~\cite{bowman2022spectral}. Given the strongly anisotropic curvature that typically arises in high dimensional potential energy surfaces, these spectral biases can substantially affect both the efficiency and the accuracy. A principled understanding of interactions between the optimizer and the loss landscape in this setting is still limited, and such an understanding is crucial for designing reproducible and physically consistent MLIP workflows.

\begin{figure}[t]
\centering
\includegraphics[width=0.95\linewidth]{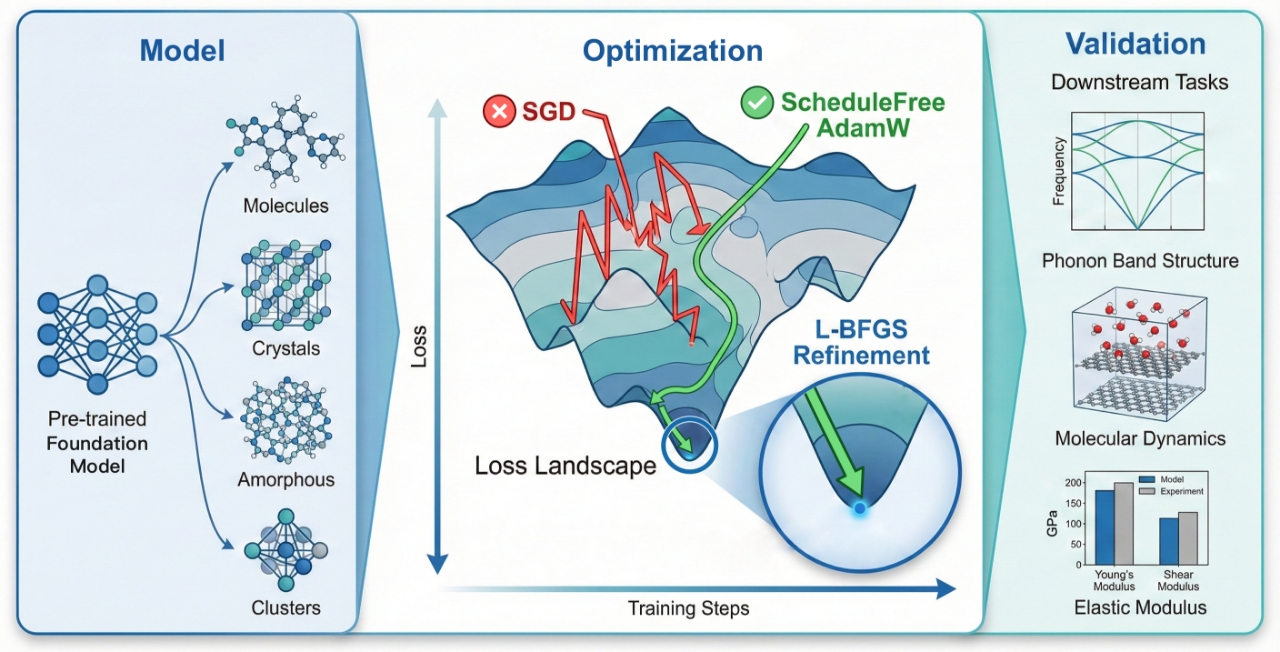}
\caption{Schematic illustration of the study design. A pretrained MACE-based foundation model is fine-tuned on inorganic, molecular, and liquid benchmarks using various first-order optimizers, optionally followed by an L-BFGS refinement stage.}
\label{fig:schema}
\end{figure}

In this study, we address this gap by combining a comprehensive empirical benchmark with a geometric analysis of optimization dynamics. We utilize a foundation model based on the MACE architecture as a common backbone to perform fine-tuning across inorganic, organic, and liquid systems. We evaluate seven representative first-order optimizers: Adam~\cite{kingma2014adam}, AdamW~\cite{loshchilov2017decoupled}, RAdam~\cite{liu2019variance}, stochastic gradient descent (SGD)\cite{loshchilov2016sgdr}, LAMB\cite{you2019large}, Ranger~\cite{tong2022calibrating}, and ScheduleFree~\cite{defazio2024road}. Our empirical analysis comprises two distinct phases. First, we characterize convergence behavior and generalization capabilities on the Silicon and 3BPA benchmarks, covering both in-distribution and out-of-distribution regimes. Second, we establish a link between the choice of optimizer and physical fidelity by evaluating elastic moduli, migration barriers, phonon spectra, and dynamical observables. We further investigate the efficacy of a second-order refinement phase using L-BFGS~\cite{liu1989limited}. Specifically, we analyze whether this post-processing step effectively sharpens the learned potential energy surface and quantify the trade-off between accuracy gains and the associated computational overhead. Finally, we interpret all results within a unified preconditioning framework. This perspective treats each optimizer as a data-dependent linear operator acting on the gradient field, allowing us to analyze the induced spectral filtering of the anisotropic loss landscape. An overview of the workflow and evaluation stages is presented in Figure~\ref{fig:schema}.

Our contributions are threefold. First, we introduce a systematic protocol for comparing optimization algorithms in the fine-tuning of atomistic foundation models that jointly evaluates conventional energy and force metrics alongside downstream physical properties. Second, we identify robust trends across optimization strategies. We find that AdamW and ScheduleFree yield superior curvature conditioning and produce smoother and more transferable potential energy surfaces compared to Adam. We provide a mechanistic explanation for these observations based on the preconditioning analysis developed in this work. Third, we delineate the specific regimes where an L-BFGS post-processing step offers considerable benefits relative to its computational cost. We demonstrate that this refinement reliably improves force accuracy and local property predictions once first-order training has reached a basin of attraction. Collectively, these results indicate that the choice of optimizer is a fundamental design variable in the development and deployment of U-MLIPs rather than a minor implementation detail.

\section{Methods}
\label{sec:methods}

This section presents the methodological framework used to analyze how optimization algorithms influence the fine-tuning behavior of atomistic foundation models. Section~\ref{sec:sub:mace} introduces the MACE architecture and the fine-tuning formulation adopted in this study, including the training objective and the gradient based update scheme. Section~\ref{sec:sub:optimizers} summarizes the representative optimization algorithms benchmarked in our work and highlights their characteristic update rules and the underlying design principles. Section~\ref{sec:sub:preconditioning} develops a unified theoretical interpretation of these algorithms from a preconditioning perspective.

\subsection{The MACE foundation model and fine-tuning framework}
\label{sec:sub:mace}

This subsection describes the pretrained foundation models used throughout this work and the fine-tuning protocol that enables a controlled comparison of optimization strategies. Although our study focuses on MACE based models, the concepts and conclusions are broadly relevant to other universal MLIPs.

\subsubsection{MACE-MP-0 and MACE-OFF foundation models.}
\label{sec:sub:mace_models}

U-MLIPs aim to provide transferable representations of atomic environments across wide chemical and structural ranges. We adopt the MACE architecture~\cite{batatia2022mace}, which combines the systematically improvable body order construction of the Atomic Cluster Expansion~\cite{ACECompleteness, DrautzACE} with an equivariant message passing network. The architecture builds tensorial features that are invariant to translation and permutation and equivariant to rotation, and combines these features through tensor products that capture correlations across multiple interaction orders.

On top of this architecture, large scale pre-training produces concrete foundation models. In the inorganic domain, we use MACE-MP-0~\cite{batatia2023foundation}, which is trained on chemically diverse structures and trajectories drawn from major materials databases and associated high throughput workflows. Its training corpus spans a wide portion of the periodic table and contains crystalline, amorphous, and defected configurations, which enables the model to learn transferable descriptors for extended solids and elemental combinations relevant to materials chemistry. Several refined variants exist, including MACE-MP-0b, MACE-MP-0b2, MACE-MP-0b3, MACE-MPA-0, and MACE-OMAT-0, each targeting improved behavior in specific physical regimes such as short range repulsion, high pressure stability, or phonon accuracy. These models are publicly available~\cite{macefoundations-github} and constitute a well characterized family of inorganic universal MLIPs.

For molecular and organic chemistry, we employ MACE-OFF-23~\cite{kovacs2025mace}, which is pretrained on multi temperature conformational ensembles, torsional scans, and diverse molecular datasets that emphasize flexibility, intramolecular rearrangements, and noncovalent interactions. This training distribution complements that of MACE-MP-0 by covering chemical motifs and energy landscapes characteristic of small and medium sized molecules. An updated version, MACE-OFF-24, has recently been released; although it is not explicitly benchmarked here, we expect that the qualitative optimizer trends reported in this work would extend to that model as well.

Together, MACE-MP-0 and MACE-OFF provide two contrasting but complementary foundation settings. The former represents extended inorganic structures and broad compositional diversity, whereas the latter captures the rich conformational variability of molecular systems. Using both models allows us to examine optimizer behavior across distinct types of loss landscapes. Throughout all experiments, we adopt the multi-head replay fine-tuning framework~\cite{batatia2025cross}. The model employs a dual-head architecture sharing a common MACE backbone. The first head serves as a replay mechanism and is supervised by a small representative subset of the original pre-training dataset to preserve general chemical knowledge and mitigate catastrophic forgetting. The second head is dedicated to the specific downstream task and is trained on the new target dataset. This design ensures that the model adapts to the specific system while maintaining the robustness of the pre-trained representation.

\subsubsection{Training framework.}
\label{sec:sub:train}

Fine-tuning adapts a pre-trained model to a specific target system by optimizing its parameters against high fidelity reference data, e.g., DFT. Let $\{(R_n,E_n,F_n)\}_{n=1}^{N_{\mathrm{conf}}}$ denote the training set, where $R_n\in\mathbb{R}^{3N_n}$ is the atomic coordinate vector of configuration $n$ with $N_n$ atoms, $E_n$ is its total energy, and $F_n\in\mathbb{R}^{3N_n}$ is the corresponding force vector obtained from electronic structure calculations. The model predicts an energy $E_\theta(R)$ and forces $F_\theta(R) = -\nabla_R E_\theta(R)$, so that energy conservation is enforced by construction. Fine-tuning proceeds by minimizing the standard joint energy and force regression objective
\begin{equation}
\label{eq:loss-ft}
\mathcal{L}(\theta)
=
\frac{1}{N_{\mathrm{conf}}}
\sum_{n=1}^{N_{\mathrm{conf}}}
\!\left[
w_E \bigl(E_\theta(R_n)-E_n\bigr)^2
+
\frac{w_F}{3N_n}\,
\bigl\|F_\theta(R_n)-F_n\bigr\|_2^2
\right],
\end{equation}
where $w_E$ and $w_F$ control the relative emphasis on matching total energies and atomic forces. In some settings, a virial or stress term is added to Eq.~\eqref{eq:loss-ft} by including an additional quadratic penalty on the predicted stress tensor, but this augmentation is conceptually identical and does not change the optimization framework used here.

To minimize Eq.~\eqref{eq:loss-ft}, parameter updates follow the iterative rule
\begin{equation}
\label{eq:update}
\theta^{(t+1)}
=
\theta^{(t)}
-
\eta_t\,P_t(\theta^{(t)})\,g_t,
\qquad
g_t=\nabla_\theta\mathcal{L}(\theta^{(t)}),
\end{equation}
where $\eta_t$ is the learning rate and $P_t$ is the preconditioning operator associated with a particular optimizer. Different choices of $P_t$ determine how gradient components are rescaled or regularized during training and therefore control how the pretrained representation adapts to the target domain, see Section~\ref{sec:sub:preconditioning}. The behavior of $P_t$ plays a central role in whether fine-tuning remains stable, whether the learned relation between energy and forces is physically smooth, and how quickly the model converges.

In the remainder of this work, we systematically evaluate how commonly used first-order optimizers instantiate $P_t$ in Eq.~\eqref{eq:update}, and we quantify how these differences influence accuracy, stability, and physical fidelity in downstream applications.

\subsection{Representative first-order optimizers}
\label{sec:sub:optimizers}

Within the preconditioning framework of Eq.~\eqref{eq:update}, the optimizers examined in this study differ mainly in how they shape gradient magnitudes during fine-tuning.

\subsubsection{Stochastic gradient descent.}
\label{sec:sgd}

Stochastic gradient descent (SGD)~\cite{amari1993backpropagation} provides the simplest instance of Eq.~\eqref{eq:update}, with the identity preconditioner $P_t = I$ so that the parameters are updated directly along the instantaneous gradient,
\begin{equation}
\theta^{(t+1)} = \theta^{(t)} - \eta_t\, g_t,
\qquad
g_t = \nabla_\theta \mathcal{L}(\theta^{(t)}).
\end{equation}
In this formulation, the effective step size is controlled solely by the global learning rate $\eta_t$, and all directions in the parameter space are treated uniformly.

On the local quadratic model introduced in Eq.~\eqref{eq:quad}, with Hessian $H$, the convergence rate of SGD is controlled by the condition number of $H$. Eigenmodes with large eigenvalues impose a strict limit on the maximal stable step size, whereas modes with small eigenvalues relax only slowly. This theory shows that an identity preconditioner is inherently inefficient when the spectrum of $H$ is highly spread~\cite{bottou2018optimization, nesterov2003introductory} .

Fine-tuning U-MLIPs typically produce a loss landscape for the joint energy and force objective with strongly anisotropic curvature. Since SGD employs a single global learning rate and does not adapt to this structure, it either advances slowly along flat directions or becomes sensitive to stiff directions. In this work, plain SGD is, therefore, used primarily as a reference method: it reflects the intrinsic conditioning of the fine-tuning problem and provides a baseline for stability and accuracy against which adaptive optimizers can be compared.

\subsubsection{Adam and its variants.}
\label{sec:adam_family}

The slow and unstable behavior of SGD in stiff loss landscapes motivates
optimizers to adapt their step sizes to local gradient statistics. In 
preconditioned update \eqref{eq:update}, these methods can be viewed as
constructing a diagonal approximation to the inverse Hessian $H^{-1}$, so that each parameter is updated with its own effective learning rate. As discussed in~\ref{sec:apd:analysis}, such diagonal preconditioning can reduce the condition number of the local Hessian and thereby accelerate convergence.

\paragraph{Adam.}

Adam~\cite{kingma2014adam} introduces elementwise adaptive scaling through
exponential moving averages of the first and second moments of the gradient,
\begin{equation}
m_t = \beta_1 m_{t-1} + (1 - \beta_1) g_t,
\qquad
v_t = \beta_2 v_{t-1} + (1 - \beta_2) g_t^2,
\end{equation}
where $\beta_1,\beta_2 \in (0,1)$ are exponential decay factors and $t$ is the
iteration index. The bias corrected estimates are defined as
\begin{equation}
\hat m_t = \frac{m_t}{1 - \beta_1^{\,t}},
\qquad
\hat v_t = \frac{v_t}{1 - \beta_2^{\,t}},
\nonumber 
\end{equation}
where $\beta_1^{\,t}$ and $\beta_2^{\,t}$ denote $\beta_1$ and $\beta_2$ raised
to the power $t$. The parameter update then reads
\begin{equation}
\theta^{(t+1)}
=
\theta^{(t)}
-
\eta_t
\frac{\hat m_t}{\sqrt{\hat v_t} + \varepsilon},
\end{equation}
where $\varepsilon > 0$ is a small constant, typically on the order of
$10^{-8}$, added elementwise to avoid division by zero and to control the
conditioning of the denominator.

This update corresponds to a diagonal preconditioner
\begin{equation}
P_t
=
\mathrm{diag}\bigl((\sqrt{\hat v_t} + \varepsilon)^{-1}\bigr),
\end{equation}
which acts as a coarse approximation to $H^{-1}$. Directions that repeatedly
exhibit large gradient magnitudes, indicative of high curvature, are damped by
large entries in $\hat v_t$, while flatter directions receive relatively
larger effective step sizes. For many MLIPs, this already yields substantially
faster and more stable fine-tuning than SGD. At the same time, noisy estimates
of the second moment on small or heterogeneous targets can perturb the balance
between energy and force errors, which motivates the Adam variants considered
below.

\paragraph{AdamW.} AdamW~\cite{loshchilov2017decoupled} modifies Adam by separating weight decay from the adaptive update,
\begin{equation}
\theta^{(t+1)}
=
\theta^{(t)}
-
\eta_t
\frac{\hat m_t}{\sqrt{\hat v_t} + \varepsilon}
-
\eta_t \lambda_{\mathrm{wd}} \theta^{(t)},
\end{equation}
where $\lambda_{\mathrm{wd}} \ge 0$ is the weight decay coefficient that controls the strength of the $L_2$ regularization on the parameters. Rather than absorbing the penalty $L_2$ in the gradient itself, AdamW applies weight decay as a separate additive term. In the preconditioning view, this preserves the diagonal scaling provided by $\hat v_t$ while applying an additional isotropic shrinkage to the parameters. Empirically, this tends to stabilize parameter norms during fine-tuning of pre-trained model and often produces smoother potential energy surfaces and better transferability than Adam.

\paragraph{RAdam.} A known weakness of Adam is its sensitivity during the first few hundred steps, when the second moment estimates are still inaccurate. The rectified Adam (RAdam) algorithm~\cite{liu2019variance} introduces a time dependent factor
$r_t \in (0,1]$ and uses
\begin{equation}
\theta^{(t+1)}
=
\theta^{(t)}
-
\eta_t r_t
\frac{\hat m_t}{\sqrt{\hat v_t} + \varepsilon}.
\end{equation}
At early iterations $r_t$ is small, and the method behaves more closely to a momentum
scheme with limited adaptivity. As the variance estimates stabilize, $r_t$
approaches one and the full Adam style scaling is recovered. This warm up of
the adaptive component reduces the risk of overly aggressive updates when the
second moment statistics are still unreliable.

\paragraph{LAMB.}

Layerwise Adaptive Moments (LAMB)~\cite{you2019large} extends the AdamW update by applying a layerwise normalization to the proposed step. In the preconditioned form of Eq.~\eqref{eq:update}, AdamW uses
\begin{equation}
P_t^{\mathrm{adam}} =
\mathrm{diag}\!\left((\sqrt{\hat v_t}+\varepsilon)^{-1}\right),
\qquad
u_t = P_t^{\mathrm{adam}}\,\hat m_t. \nonumber 
\end{equation}
LAMB rescales this raw update $u_t$ on each layer according to the ratio
between the parameter norm and the update norm,
\begin{equation}
\Delta\theta_t
=
\frac{\|\theta_t\|_2}{\|u_t\|_2}\,u_t,
\end{equation}
so that the effective step length for a layer is aligned with its parameter
scale. This layerwise trust ratio enforces that layers with very different
norms move with comparable relative step sizes. For deep equivariant
architectures whose layers differ significantly in scale, LAMB equalizes
effective learning rates across depth and stabilizes fine-tuning without the
need for layer specific hyperparameter tuning~\cite{liu2021learning}.

\paragraph{Ranger.}

Ranger~\cite{tong2022calibrating} combines RAdam with Lookahead averaging~\cite{zhang2019lookahead}.
The fast weights follow the rectified Adam update with the corresponding
diagonal preconditioner $P_t^{\mathrm{radam}}$ in Eq.~\eqref{eq:update}. In
parallel, a second set of parameters, denoted by $\theta_{\mathrm{slow}}$,
tracks a smoothed version of the fast trajectory. In every $k$ inner steps, the
slow weights are updated by interpolating towards the current fast weights,
\begin{equation}
\theta_{\mathrm{slow}}
\leftarrow
\theta_{\mathrm{slow}}
+
\alpha\bigl(\theta_{\mathrm{fast}}-\theta_{\mathrm{slow}}\bigr),
\end{equation}
after which $\theta_{\mathrm{fast}}$ is reset to $\theta_{\mathrm{slow}}$.
The fast weights therefore explore the local landscape under
$P_t^{\mathrm{radam}}$, while the slow weights advance on a coarser time scale
that averages out short term fluctuations. In the preconditioning view,
$P_t^{\mathrm{radam}}$ governs the local spectral geometry, and Lookahead adds a
second slower timescale that smooths oscillations. This two timescale
mechanism improves robustness when batch sizes are small or gradient variance
is high.

\paragraph{ScheduleFree.}

ScheduleFree~\cite{defazio2024road} retains the AdamW diagonal preconditioner
but introduces a global scale factor $s_t$ inferred from gradient statistics.
The update remains in the form of Eq.~\eqref{eq:update} with
\begin{equation}
P_t^{\mathrm{sf}}
=
\mathrm{diag}\!\left(
s_t(\sqrt{\hat v_t}+\varepsilon)^{-1}
\right).
\end{equation}
In the analysis of Section~\ref{sec:sub:preconditioning}, this can be viewed as
a global contraction control mechanism. As long as $s_t$ stays within a
suitable range, the dominant eigenvalues of the iteration matrix remain within
a stable contraction interval throughout training. This reduces the need for
manually designed learning rate schedules and decreases hyperparameter
sensitivity in transfer learning~\cite{defazio2024road}.

\subsubsection{A second-order optimizer: L-BFGS.}
\label{sec:lbfgs}

The previous subsections introduced first-order optimizers that control how the parameters evolve over the entire fine-tuning process. However, the local geometry of the loss landscape in the neighborhood of a minimizer is inherently second-order (cf.~Section~\ref{sec:sub:preconditioning}). In the late stages of training, Adam type diagonal preconditioners primarily rescale gradient components and cannot fully capture curvature couplings between parameters. This limitation often leads to slow
progress along flat directions and small residual inconsistencies between energy and force predictions.

To refine the local minimizers produced by first-order optimization, we apply a limited memory BFGS (L-BFGS)~\cite{liu1989limited} procedure to the objective $\mathcal{L}(\theta)$. L-BFGS constructs an approximation \(P_k^{\mathrm{lbfgs}}\) to the inverse Hessian of
$\mathcal{L}$ at iteration \(k\). The parameters are updated according to
\begin{equation}
\label{eq:lbfgs-update-main}
\theta_{k+1}
=
\theta_k
-
P_k^{\mathrm{lbfgs}} g_k,
\qquad
g_k = \nabla_\theta \mathcal{L}(\theta_k),
\end{equation}
where \(P_k^{\mathrm{lbfgs}}\) is a symmetric positive definite matrix that incorporates second-order information through curvature pairs. 
The action of \(P_k^{\mathrm{lbfgs}}\) on \(g_k\) can be interpreted as a dense
preconditioning operation that couples coordinates and adapts the step sizes to the
local curvature. On smooth quadratic objectives this yields a convergence rate
close to that of exact Newton updates, but at a fraction of the computational
cost of forming or inverting the full Hessian. The algorithmic details of the
L-BFGS implementation used in this work are given in~\ref{sec:apd:lbfgs}.

From the preconditioning point of view, L-BFGS complements the diagonal scaling
used by Adam type optimizers. The first-order stage reduces large scale
anisotropy and moves the parameters into a suitable basin of attraction, while
the quasi Newton refinement incorporates local curvature information to
accelerate the final phase of convergence. For the U-MLIPs considered
here, this combination yields improved force accuracy in held out configurations, and improved stability in downstream molecular dynamics simulations, as demonstrated in Section~\ref{sec:results}.

\subsection{Optimizer induced preconditioning perspective}
\label{sec:sub:preconditioning}

The update rule in Eq.~\eqref{eq:update} can be rigorously formulated as a preconditioned gradient iteration. In this framework, the optimizer constructs a data-dependent linear operator \(P_t(\theta^{(t)})\) that modifies the gradient vector to account for the local curvature of the objective function. This operator effectively defines a variable metric on the parameter space, which allows the optimization trajectory to adapt to the anisotropy of the loss landscape. This perspective unifies the various algorithms presented in Section~\ref{sec:sub:optimizers} and provides the necessary mechanistic insight to interpret the convergence and stability patterns reported in Section~\ref{sec:results}. A comprehensive spectral analysis of the eigenvalue distributions resulting from diagonal preconditioning is provided in~\ref{sec:apd:analysis}.

\subsubsection{Local spectral analysis.}

We analyze the behavior of local convergence through a quadratic expansion of loss \(\mathcal{L}(\theta)\) around a local minimizer \(\theta^{*}\):
\begin{equation}
\label{eq:quad}
\mathcal{L}(\theta)
\simeq
\mathcal{L}(\theta^{*})
+
\tfrac12(\theta-\theta^{*})^{\mathsf T}
H
(\theta-\theta^{*}),
\qquad
H=\nabla_\theta^2 \mathcal{L}(\theta^{*}),
\end{equation}
where \(H\) is the Hessian matrix evaluated at \(\theta^{*}\). Defining the parameter error vector \(\delta_t=\theta^{(t)}-\theta^{*}\) and approximating the gradient as \(g_t \approx H\delta_t\), the update rule in Eq.~\eqref{eq:update} becomes a linear recurrence relation:
\begin{equation}
\label{eq:lin-dyn}
\delta_{t+1}
=
\left(I-\eta_t P_t H\right)\delta_t.
\end{equation}
The matrix \(M_t = I-\eta_t P_t H\) governs the contraction of the error vector. The stability and convergence speed of the optimization are determined by the spectral radius and the eigenvalue distribution of \(M_t\).

In the case of SGD, where \(P_t = I\), the contraction dynamics is dictated entirely by the Hessian spectrum \(H\). The convergence rate is limited by the condition number \(\kappa(H) = \lambda_{\max}/\lambda_{\min}\), where \(\lambda_{\max}\) and \(\lambda_{\min}\) are the largest and smallest eigenvalues of \(H\). For U-MLIPs, the physical coexistence of stiff high-frequency modes and soft low-frequency interactions typically results in a highly anisotropic Hessian with large \(\kappa(H)\). Consequently, SGD is forced to employ small step sizes to ensure stability in stiff directions, which leads to inefficient stagnation along flat directions.

Preconditioning fundamentally alters this spectral structure by inserting \(P_t\) into the gradient update. In the Hessian eigenbasis, \(P_t\) rescales the contributions of the curvature, which replaces the raw eigenvalues \(\lambda_i\) with the effective eigenvalues of the product \(P_t H\). An effective preconditioner compresses the spectrum of \(P_t H\) and reduces the effective condition number. This spectral compression mitigates the disparity between fast and slow modes, which allows the use of larger learning rates and ensures uniform convergence across the parameter space.~\ref{sec:apd:analysis} formalizes these spectral properties for diagonal preconditioners and quantifies how different scaling laws influence the effective condition number.

\subsubsection{Practical implications for MLIP optimizers.}

In practice, modern optimizers implement preconditioning through diagonal or
layer structures in \(P_t\), combined with mechanisms that control the
overall step scale. For the optimizers studied here, the dominant effects can
be summarized as follows.

% \begin{table}[t]
% \setlength{\belowcaptionskip}{6pt}  % 只作用于当前环境
% \centering
% \caption{Preconditioning behavior of the optimizers considered in this work.}
% \label{tab:optim_precond}
% \begin{scriptsize}
% \renewcommand{\arraystretch}{1.3}
% \begin{tabular}{@{}l p{0.22\linewidth} p{0.60\linewidth}@{}}
% \toprule
% {\bf Optimizer} & {\bf Structure of \(P_t\)} & {\bf Qualitative effect on fine-tuning} \\ 
% \midrule
% SGD &
% identity matrix &
% no curvature normalization; progress limited by \(\kappa(H)\) \\

% Adam &
% diagonal from second moment \(v_t\) &
% strong elementwise flattening of stiff directions; faster convergence but potentially noisy \\

% AdamW &
% Adam diagonal plus weight decay &
% better norm control and smoother potential energy surfaces in transfer \\

% RAdam &
% time dependent scaling of Adam step &
% safer early iterations while variance estimates are unreliable \\

% LAMB &
% layerwise normalized AdamW step &
% balanced relative step sizes across layers in deep equivariant stacks \\

% Ranger &
% RAdam with Lookahead averaging &
% trajectory smoothing and improved robustness under high gradient variance \\

% ScheduleFree &
% globally scaled AdamW preconditioner &
% automatic control of global gain; reduced reliance on manual learning rate schedules \\

% L-BFGS &
% low rank dense inverse Hessian approximation &
% near isotropy on a subspace; improves terminal forces and energy smoothness \\
% \bottomrule
% \end{tabular}
% \end{scriptsize}
% \end{table}

Adam, AdamW, and RAdam primarily perform diagonal curvature normalization by
using running estimates of gradient moments to approximate a diagonal inverse
Hessian. This flattens the spectrum of \(H\) at the level of individual
parameters and improves conditioning compared with SGD. LAMB and Ranger add
additional control of the trajectory through layerwise normalization and
two time scale averaging, which is particularly beneficial for deep equivariant
architectures with heterogeneous layer scales. ScheduleFree uses a global scale
factor on top of AdamW style preconditioning to keep the iteration matrix
\(M_t\) within a stable contraction regime over the course of training. Finally, the L-BFGS refinement stage introduces a complementary low rank dense
correction to the diagonal preconditioning provided by Adam type optimizers.
This captures curvature couplings between parameters near the attained minimum
and removes residual anisotropy that limits the terminal phase of fine-tuning.

For the U-MLIPs considered in this work, these mechanisms translate
directly into differences in convergence speed, stability, and the balance
between energy and force accuracy. The numerical results in
Section~\ref{sec:results} show that optimizers with stronger effective
preconditioning tend to produce smoother potential energy surfaces and a more robust transfer across the thermodynamic and structural regimes.

\section{Numerical Results}
\label{sec:results}

In this section, we empirically assess how different optimization strategies
affect the fine-tuning of atomistic foundation models. 
Section~\ref{sec:sub:benchmark} presents a controlled benchmark of first-order
optimizers on representative organic and inorganic systems, focusing on energy
and force accuracy as well as out of distribution generalization.
Section~\ref{sec:sub:second_order} then examines the impact of an additional
second-order refinement stage on the accuracy of the property level and the stability of the molecular dynamics.

\subsection{First-order optimizer comparison on organic and inorganic fine-tuning}
\label{sec:sub:benchmark}

\begin{figure}[t]
    \centering
    \includegraphics[width=0.9\linewidth]{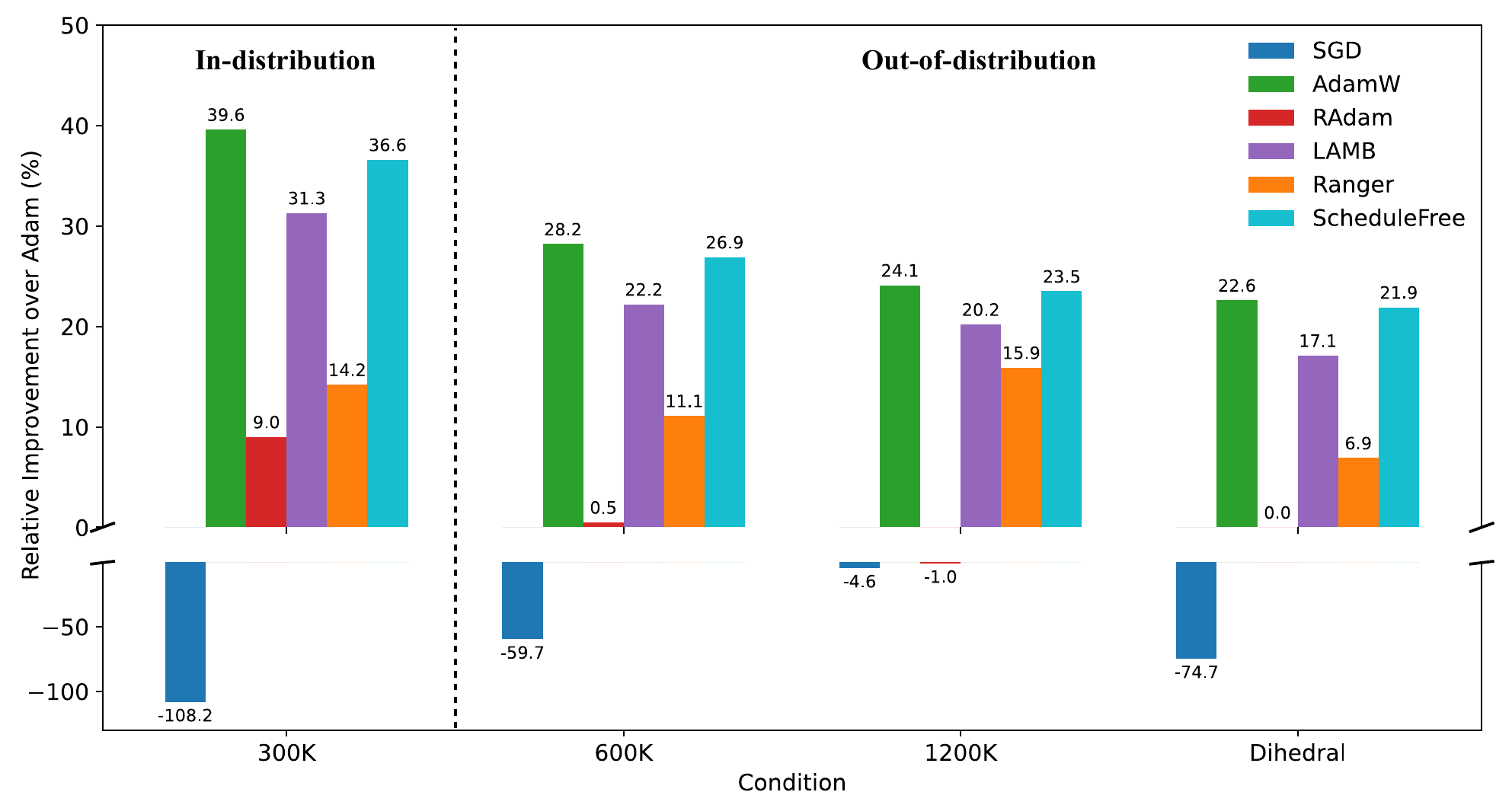}
    \caption{Force RMSE relative improvement compared with Adam across four test conditions. Negative values indicate worse performance than Adam.}
    \label{fig:rel_improve}
\end{figure}

We begin with a comparison of seven optimizers in two representative
universal MLIP scenarios: the organic 3BPA benchmark based on the MACE-OFF-23
pretrained model and the inorganic silicon benchmark based on the MACE-MPA-0
pretrained model. Tables~\ref{tab:benchmarks_3bpa} and
\ref{tab:benchmarks_si} report the root mean square errors for the energies
(E, meV/atom) and forces (F, meV/\AA) in these two systems.
For 3BPA, the model is fine tuned on configurations at 300 K, while the test set covers both in-distribution configurations at 300 K and out-of-distribution configurations at 600 K, 1200 K, and along a dihedral scan~\cite{batatia2022mace}. The silicon benchmark similarly probes generalization to structures not present in the training set, including stacking faults and amorphous (a-Si) phases~\cite{bartok2018machine}.
Figure~\ref{fig:rel_improve} visualizes the relative improvements over Adam on 3BPA. All experiments are repeated with three random seeds, and we report mean and standard deviation across seeds in order to quantify statistical variability.

\begin{table*}[h]
\caption{RMSE of energy (E, meV/atom) and force (F, meV/\AA) on the 3BPA dataset. 
The training set is collected at 300 K. Standard deviations are computed over three runs and shown in brackets. 
The best two results of each conditions are in bold.}
\label{tab:benchmarks_3bpa}
\begin{center}
\begin{scriptsize}
\renewcommand{\arraystretch}{1.4} % 调大行距
\begin{tabular}{lccccccc}
\toprule
{\bf Condition} & {\bf SGD} & {\bf Adam} & {\bf AdamW} & {\bf RAdam} & {\bf LAMB} & {\bf Ranger} & {\bf ScheduleFree} \\
\midrule
\multirow{2}{*}{300K} 
& 0.7 (0.01) & 0.4 (0.12) & {\bf 0.2 (0.01)} & 0.6 (0.01) & {\bf 0.2 (0.01)} & 0.3 (0.01) & {\bf 0.2 (0.01)} \\
 & 27.9 (0.06) & 13.4 (1.70) & {\bf 8.1 (0.02)} & 12.2 (0.06) & 9.2 (0.02) & 11.5 (0.02) & {\bf 8.5 (0.02)} \\
\hline
\multirow{2}{*}{600K} 
& 1.0 (0.01) & 0.5 (0.06) & {\bf 0.4 (0.06)} & {\bf 0.4 (0.01)} & 0.5 (0.21) & {\bf 0.4 (0.06)} & {\bf 0.3 (0.01)} \\
& 34.5 (0.02) & 21.6 (0.21) & {\bf 15.5 (0.06)} & 21.5 (0.21) & 16.8 (0.06) & 19.2 (0.02) & {\bf 15.8 (0.06)} \\
\hline
\multirow{2}{*}{1200K} 
& 1.4 (0.01) & 0.9 (0.01) & {\bf 0.7 (0.01)} & 0.9 (0.01) & 0.8 (0.12) & {\bf 0.7 (0.06)} & {\bf 0.6 (0.06)} \\
& 52.5 (0.01) & 50.2 (0.49) & {\bf 38.1 (0.12)} & 50.7 (0.85) & 40.1 (0.23) & 42.2 (0.10) & {\bf 38.4 (0.15)} \\
\hline
\multirow{2}{*}{Dihedral} 
& 1.1 (0.01) & 0.6 (0.01) & 0.4 (0.06) & {\bf 0.3 (0.01)} & 0.4 (0.23) & {\bf 0.3 (0.06)} & {\bf 0.2 (0.06)} \\
& 25.5 (0.06) & 14.6 (0.15) & {\bf 11.3 (0.10)} & 14.6 (0.10) & 12.1 (0.06) & 13.6 (0.15) & {\bf 11.4 (0.06)} \\
\bottomrule
\end{tabular}
\end{scriptsize}
\end{center}
\end{table*}

Across both domains, the results exhibit a clear and consistent pattern that aligns with the preconditioning analysis in Section~\ref{sec:sub:preconditioning}.
Default Adam is not the top-performing optimizer: although its second-moment estimates provide strong early conditioning, variance accumulation in the denominator often yields late-stage oscillations and slightly suboptimal force RMSE.
In contrast, AdamW, ScheduleFree, and LAMB outperform Adam on nearly all metrics, typically reducing force RMSE by 5–15\% under identical training budgets.
These improvements are especially pronounced in the 3BPA benchmark, whose complex torsional landscape produces a highly anisotropic and stiff loss surface.
Optimizers equipped with more stable diagonal scaling (AdamW, ScheduleFree) or layer-wise normalization (LAMB) are better matched to this curvature structure, leading to faster convergence and better generalization.
The inorganic silicon benchmark exhibits a similar trend, indicating that these gains are not limited to molecular systems but also transfer to large-scale inorganic MLIPs.

SGD performs the worst in all settings, converging slowly and often stalling, which is consistent with the absence of curvature-aware scaling in its updates.
RAdam and Ranger offer intermediate behavior: RAdam improves early-phase stability by rectifying unreliable second-moment estimates, while Ranger further smooths the optimization trajectory via Lookahead averaging, producing some of the cleanest convergence curves but not consistently the lowest RMSE.

\begin{table*}[h]
\caption{RMSE of energy (E, meV/atom) and force (F, meV/\AA) on the Silicon dataset. Standard deviations are computed over three runs and shown in brackets. The best results are in bold.}
\label{tab:benchmarks_si}
\begin{center}
\begin{scriptsize}
\renewcommand{\arraystretch}{1.3} % 调大行距
\begin{tabular}{lccccccc}
\toprule
{\bf Condition} & {\bf SGD} & {\bf Adam} & {\bf AdamW} & {\bf RAdam} & {\bf LAMB} & {\bf Ranger} & {\bf ScheduleFree} \\
\midrule
\multirow{2}{*}{SFs} 
& 5.1 (0.01) & 2.1 (0.46) & 2.3 (0.17) & {\bf 1.9 (0.20)} & {\bf 1.8 (0.21)}
 & 2.3 (0.10)
 & 2.1 (0.09) \\
& 31.7 (0.10) & 31.8 (2.27) & {\bf 30.1 (3.61)} & 31.5 (1.22) & {\bf 30.3 (1.42)}
 & 39.0 (1.04)
  & 34.7 (1.32) \\
\hline
\multirow{2}{*}{a-Si} 
& 8.0 (0.65) & 7.7 (0.82) & {\bf 7.5 (0.15)} & 8.0 (0.23) & 7.6 (0.36) & 8.4 (0.51)  & {\bf 7.1 (0.31)} \\
& 80.7 (0.20) & 66.1 (0.45) & 66.2 (0.90) & 66.3 (0.70) & {\bf 64.3 (0.30)} & 65.0 (0.44)  & {\bf 64.0 (0.50)} \\
\bottomrule
\end{tabular}
\end{scriptsize}
\end{center}
\end{table*}

The same qualitative conclusions hold for the inorganic Silicon benchmark. Although the Si landscape differs markedly from the 3BPA molecular system, the advantage of well-behaved diagonal preconditioners remains robust: AdamW, ScheduleFree, LAMB, and RAdam all outperform Adam, while SGD lags far behind.
This consistency across-domains suggests that optimizer-induced spectral flattening quantified in \ref{sec:apd:analysis} is a central determinant of fine-tuning performance for universal MLIPs.

Finally, we report the study of ablation on the learning rate in~\ref{sec:apd:ablation}. Although the absolute RMSE values vary across temperatures, targets, and learning rate choices, the relative ranking of optimizers remains stable. Methods with strong diagonal or layer wise preconditioning, in particular AdamW and ScheduleFree, consistently achieve the best transfer performance across all energy and force evaluations and remain robust to variations in the learning rate, whereas unconditioned first order methods such as SGD remain ineffective. The remaining adaptive methods, including Adam, RAdam, LAMB, and Ranger, occupy an intermediate regime and do not close the gap between AdamW and ScheduleFree. A wall clock Pareto frontier over RMSE and training time is given in Figure~\ref{fig:pareto}, which further indicates that these accuracy gains do not incur a significant computational overhead.

\subsection{Impact of second-order refinement: from RMSE to physical properties}
\label{sec:sub:second_order}

Although the first-order optimizers analyzed in the previous section demonstrate robust global convergence, particularly AdamW and ScheduleFree, the terminal phase of fine-tuning presents a distinct challenge. In the vicinity of a local minimum, the loss landscape frequently exhibits ill-conditioned curvature. To investigate the limits of attainable accuracy, we apply a brief second-order refinement stage using the limited-memory BFGS (L-BFGS) algorithm. The L-BFGS refinement is initialized from the terminal solutions of the Adam, AdamW, and ScheduleFree baselines and proceeds for a short duration of 20–40 steps, adapted to the specific conditioning of each system.

\subsubsection{Accuracy trends across distinct energy landscapes.}
\label{sec:sub:sub:rmse}

To assess how optimizer choice and second-order refinement behave across
qualitatively different loss landscapes, we extend our benchmark to five
systems that span a range of configurational and chemical complexity:
(i) body centered cubic Mo, a prototypical metallic crystal with a relatively
smooth potential energy surface;
(ii) monolayer MoS$_2$, a transition metal dichalcogenide dominated by
approximately harmonic lattice vibrations;
(iii) Li$_x$FePO$_4$ (LFP), a battery cathode material with pronounced phase
coexistence and compositional disorder;
(iv) ZnO/GaN, a heterostructure with solid-solid interfaces and a rugged 
multiwell landscape;
and (v) a graphene-water interface, representing a solid-liquid boundary with
heterogeneous force scales.

\begin{table*}[t]
\caption{Effect of L-BFGS refinement on force RMSE (meV/\AA) across five benchmark systems. Base reports the error of the first-order optimizer, and +L-BFGS reports the error after an additional second-order refinement starting from the corresponding base model. Bold entries indicate cases where the L-BFGS refinement reduces the force RMSE relative to the base model.}
\label{tab:lbfgs_rmse}
\begin{center}
\begin{scriptsize}
\renewcommand{\arraystretch}{1.3}
\setlength{\tabcolsep}{4mm}
\begin{tabular}{lcccccc}
\toprule
\multirow{2}{*}{\textbf{System}} & \multicolumn{2}{c}{\textbf{Adam}} & \multicolumn{2}{c}{\textbf{AdamW}} & \multicolumn{2}{c}{\textbf{ScheduleFree}} \\
\cmidrule(lr){2-3} \cmidrule(lr){4-5} \cmidrule(lr){6-7}
 & Base & +L-BFGS & Base & +L-BFGS & Base & +L-BFGS \\
\midrule
\textbf{Mo} (BCC) & 32.1 & 32.5 & 28.6 & 28.6 & 23.5 & 23.5 \\ 
\textbf{MoS$_2$} (2D) & 3.1 & {\bf 2.7} & 2.2 & 2.2 & 3.9 & {\bf 2.8} \\ 
\textbf{LFP} (Cathode) & 66.8 & {\bf 66.5} & 65.4 & {\bf 65.3} & 66.2 & {\bf 66.1} \\ 
% \textbf{HECN} (High-Entropy) & 193.6 & 194.6 & 194.9 & 195.3 & 193.4 &  194.0\\ 
\textbf{ZnO/GaN}  & 9.2 & {\bf 9.1} & 7.5 & 7.5 & 9.4 &  9.4\\ 
\textbf{Gr-Water} (Interface) & 12.2 & 10.2 & 10.4 & {\bf 10.1} & 11.5 & {\bf 10.7} \\ 
\bottomrule
\end{tabular}
\end{scriptsize}
\end{center}
\end{table*}

The force RMSE data in Table~\ref{tab:lbfgs_rmse} demonstrate that the utility of second-order refinement is determined by the spectral structure of the underlying physics. In structurally homogeneous systems, such as bulk Mo and monolayer MoS$_2$, adaptive first-order methods saturate the attainable accuracy. The invariance of the error metrics under L-BFGS post-processing indicates that the local potential energy surface in these regimes is well-conditioned and adequately resolved by diagonal preconditioning.

However, systems exhibiting interfacial heterogeneity or high configurational entropy, exemplified by the Graphene-Water interface (cf.~Section~\ref{sec:sub:sub:qoi}), require quasi-Newton refinement to escape stagnation. These regimes are characterized by a pronounced stiff-soft mode disparity, where rigid covalent networks couple to fluxional solvent degrees of freedom, creating a highly anisotropic loss landscape. Although diagonal optimizers fail to capture the off-diagonal curvature correlations necessary to navigate this disparity, the low-rank inverse Hessian approximation in L-BFGS effectively rescales the coupled modes. This allows the optimizer to resolve fine-scale force variations that remain inaccessible to first-order methods.

\subsubsection{Validation on physical quantities of interest.}
\label{sec:sub:sub:qoi}

Standard error metrics provide a necessary but insufficient measure of model quality, as they often mask local topological defects in the energy landscape that govern macroscopic behavior. To assess whether refined models possess true physical predictive power, we move beyond pointwise accuracy to examine three critical regimes: static mechanical derivatives (Mo), harmonic vibrational eigenmodes (MoS$_2$), and phase-space stability at heterogeneous interfaces (graphene-water).

\paragraph{Static mechanics and energy landscapes (Mo).}

\begin{figure}[t]
    \centering
    \includegraphics[width=0.8\linewidth]{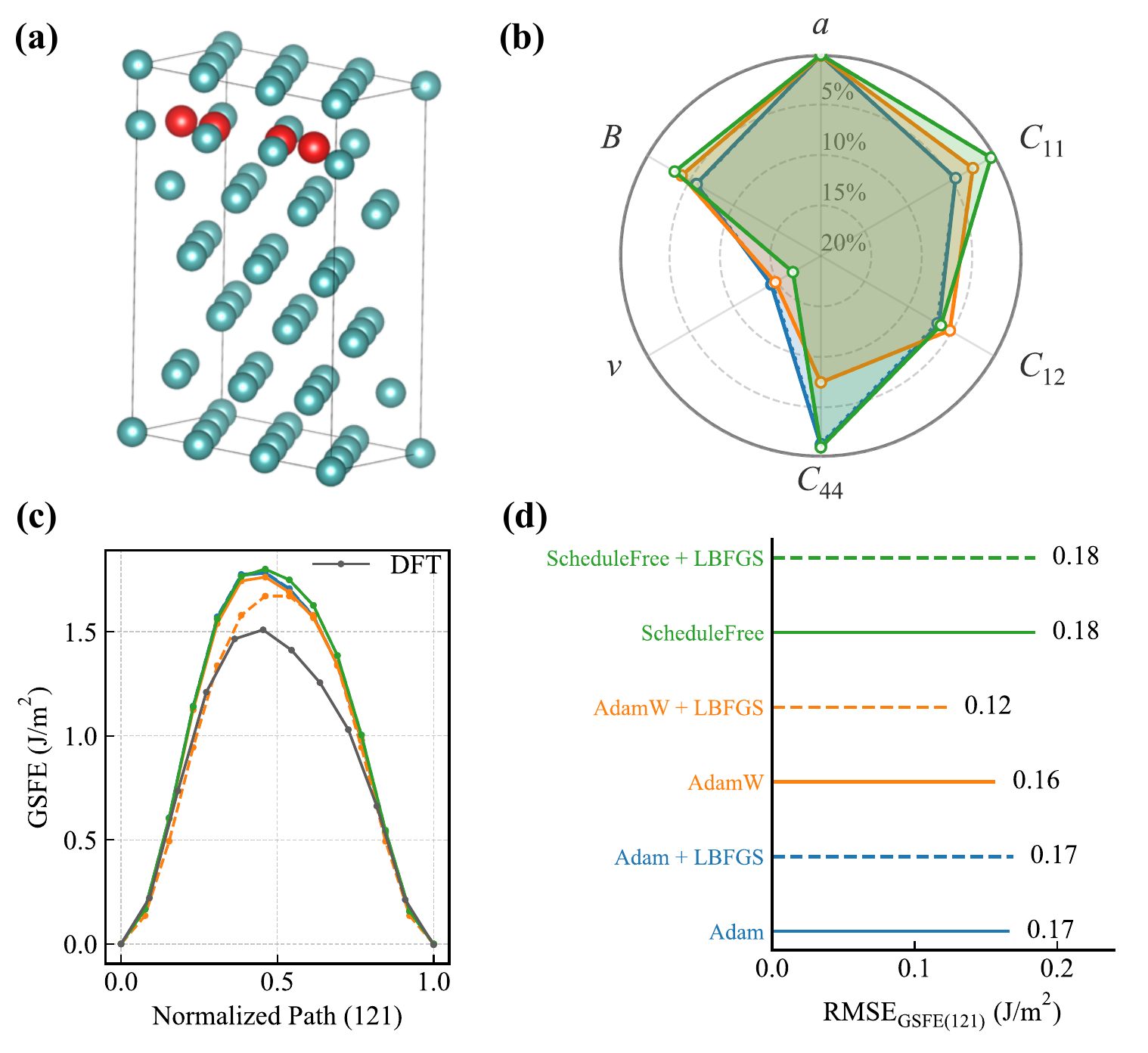}
    \caption{Static mechanical properties and GSFE for BCC Mo. {\bf (a)} Atomic structure of Mo with a dislocation, where the defect core is highlighted in red.
    {\bf (b)} Relative errors of lattice constant, elastic constants \(C_{11}\),
    \(C_{12}\), \(C_{44}\), bulk modulus \(B\), and Poisson ratio \(\nu\)
    for Adam, AdamW, and ScheduleFree, with and without L-BFGS refinement.
    {\bf (c)} GSFE along the \(\langle 121\rangle\) slip path compared with DFT.
    {\bf (d)} GSFE RMSE.}
    \label{fig:mo}
\end{figure}

We first assess the effect of the optimization choice and L-BFGS refinement on the
static mechanical response of body centered cubic Mo, whose atomic structure with defects is shown in Figure~\ref{fig:mo}(a). This analysis probes both the
harmonic regime, through the lattice constant, elastic constants, bulk modulus
, and Poisson ratio, and a moderately anharmonic regime, through the generalized
stacking fault energy (GSFE) along the \(\langle 121\rangle\) slip path~\cite{naghdi2024neural}. 

Figure~\ref{fig:mo}(b) shows the relative errors of the elastic observables
with respect to DFT. The three adaptive first-order optimizers already produce
very similar values, with ScheduleFree slightly closer to the DFT reference for
Poisson ratio and bulk modulus than Adam and AdamW. Adding an L-BFGS refinement stage leaves the entire radar profile essentially unchanged: the points corresponding to the refined models lie on top of their first order
counterparts within the plotting resolution. In particular, neither the bulk
modulus nor the elastic anisotropy indicators exhibit a systematic shift after
refinement. This insensitivity indicates that the first-order optimizers have
already located a stationary point of the elastic energy well with negligible
residual stress and that the remaining curvature anisotropy in this region is
too weak for the second-order correction to produce a measurable effect on
linear elastic response.

The GSFE results in Figure~\ref{fig:mo}(c) and (d) lead to a consistent
conclusion. The energy barriers along the \(\langle 121\rangle\) path predicted
by AdamW and ScheduleFree closely follow the DFT curve, and the refined L-BFGS 
models track the corresponding baselines almost perfectly throughout the 
slip path. The GSFE root mean square errors change by at most a few hundredths
of \(\,\mathrm{J/m^{-2}}\) after refinement. In other words,
the path dependent energy landscape is already well converged by the adaptive
first-order training, and the limiting factor is the expressive power of the
potential and the information content of the training data rather than the
local conditioning of the optimizer. For this mono elemental bulk crystal,
where the potential energy surface is relatively smooth and close to harmonic
around the equilibrium structure, diagonal preconditioning suffices to reach a
high quality minimum, and an additional second-order refinement brings no
practical benefit for static mechanical properties.

\paragraph{Vibrational fidelity and Hessian quality (MoS$_2$).}

\begin{figure}[t]
    \centering
    \includegraphics[width=0.8\linewidth]{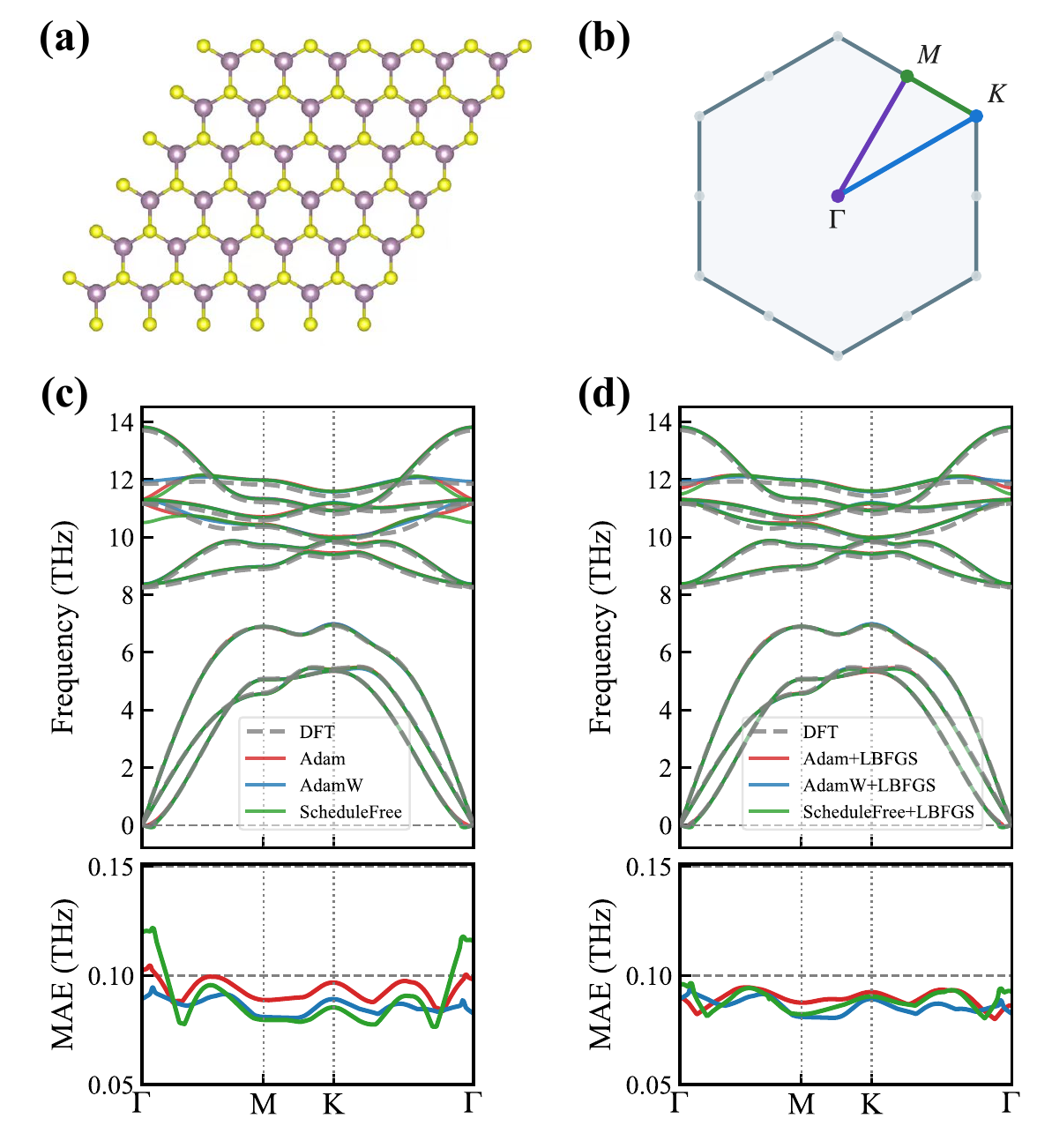}
    \caption{Phonon dispersions of monolayer MoS$_2$. 
    {\bf (a)} Relaxed MoS$_2$ monolayer. 
    {\bf (b)} First Brillouin zone and high symmetry path $\Gamma\text{--}M\text{--}K\text{--}\Gamma$. 
    {\bf (c)} Phonon spectra (top) and pointwise mean absolute error (MAE, bottom) along $\Gamma\text{--}M\text{--}K\text{--}\Gamma$ for models trained with different first-order optimizers without refinement. 
    Dashed lines denote the DFT reference.  
    {\bf (d)} Same quantities after the L-BFGS refinement.}
    \label{fig:phonons}
\end{figure}

The phonon dispersions in monolayer MoS$_2$ offer a sensitive diagnostic of the quality of learned Hessian, since the phonon frequencies are determined by its eigenvalues. Figure~\ref{fig:phonons}(a) shows the relaxed MoS$_2$ structure~\cite{mortazavi2020exploring} and Figure~\ref{fig:phonons}(b) illustrates the hexagonal Brillouin zone and the high symmetry path $\Gamma\text{--}M\text{--}K\text{--}\Gamma$ along which the dispersions are computed.

First-principles calculations were performed using VASP \cite{kresse1996efficient} with the PAW-PBE functional \cite{perdew1996generalized} and a cutoff of 500 eV. The monolayer \ch{MoS2} structure was fully relaxed (force convergence less than $0.01$ eV/Å) with a 14 Å vacuum layer. Phonon calculations were performed using Phonopy \cite{togo2015first} on a $6 \times 6 \times 1$ supercell. The second-order force constants were evaluated using two distinct methods: (i) the DFPT method \cite{gajdos2006linear} in VASP with a tightened energy convergence of $10^{-8}$ eV, and (ii) the MACE fine-tuned models.

Consistent with the low force RMSEs reported in Section~\ref{sec:sub:sub:rmse}, the phonon dispersions obtained from first-order optimization largely reproduce the DFT reference across the Brillouin zone (Figure~\ref{fig:phonons}(c)). However, a persistent artifact remains in the long-wavelength limit: the acoustic branches near the $\Gamma$-point exhibit spurious softening. This behavior signals a residual violation of the acoustic sum rules, corresponding to directions in the Hessian spectrum with near-zero or slightly negative curvature. Although these spectral defects are statistically negligible within the aggregate force loss, they introduce frequency errors on the order of \(10^{-1}\,\mathrm{THz}\) and compromise the description of thermodynamic stability.

The additional L-BFGS refinement stage corrects these deficiencies. As shown in Figure~\ref{fig:phonons}(d), a relatively small number of L-BFGS steps is sufficient to align all acoustic branches with the DFT reference and to remove the residual soft modes at $\Gamma$, indicating that the dominant negative or near zero eigenvalues have been eliminated. The corresponding pointwise frequency errors in the lower panel of Figure~\ref{fig:phonons}(d) are reduced and become nearly indistinguishable between different initial optimizers, with mean absolute errors well below \(10^{-1}\,\mathrm{THz}\). This behavior demonstrates that the refinement stage produces a well conditioned and almost optimizer independent Hessian that satisfies the acoustic sum rules and yields phonon spectra in very close agreement with the DFT reference across the entire Brillouin zone.

\paragraph{Dynamical stability at heterogeneous interfaces (Graphene--Water).}

The graphene–water interface challenges dynamical stability by coupling a stiff covalent lattice to a soft, fluxional hydrogen-bond network. This physical disparity generates a vast spectral gap that extends from high-frequency C–C stretches to slow diffusive modes. The resulting scale separation produces a highly anisotropic Hessian which standard optimizers fail to equilibrate efficiently.

Molecular dynamics simulations were performed using the LAMMPS package~\cite{thompson2022lammps} driven by MACE fine-tuned models. The equations of motion were integrated with a timestep of 0.5 fs. Following geometry optimization, the system was gradually heated from 10 K to 298 K over 10 ps. Subsequently, a production run was conducted in the canonical (NVT) ensemble at 298 K for 90 ps using the Nosé-Hoover thermostat. All computations were accelerated using a single NVIDIA A800 GPU.

\begin{figure}[t]
    \centering
    \includegraphics[width=0.8\linewidth]{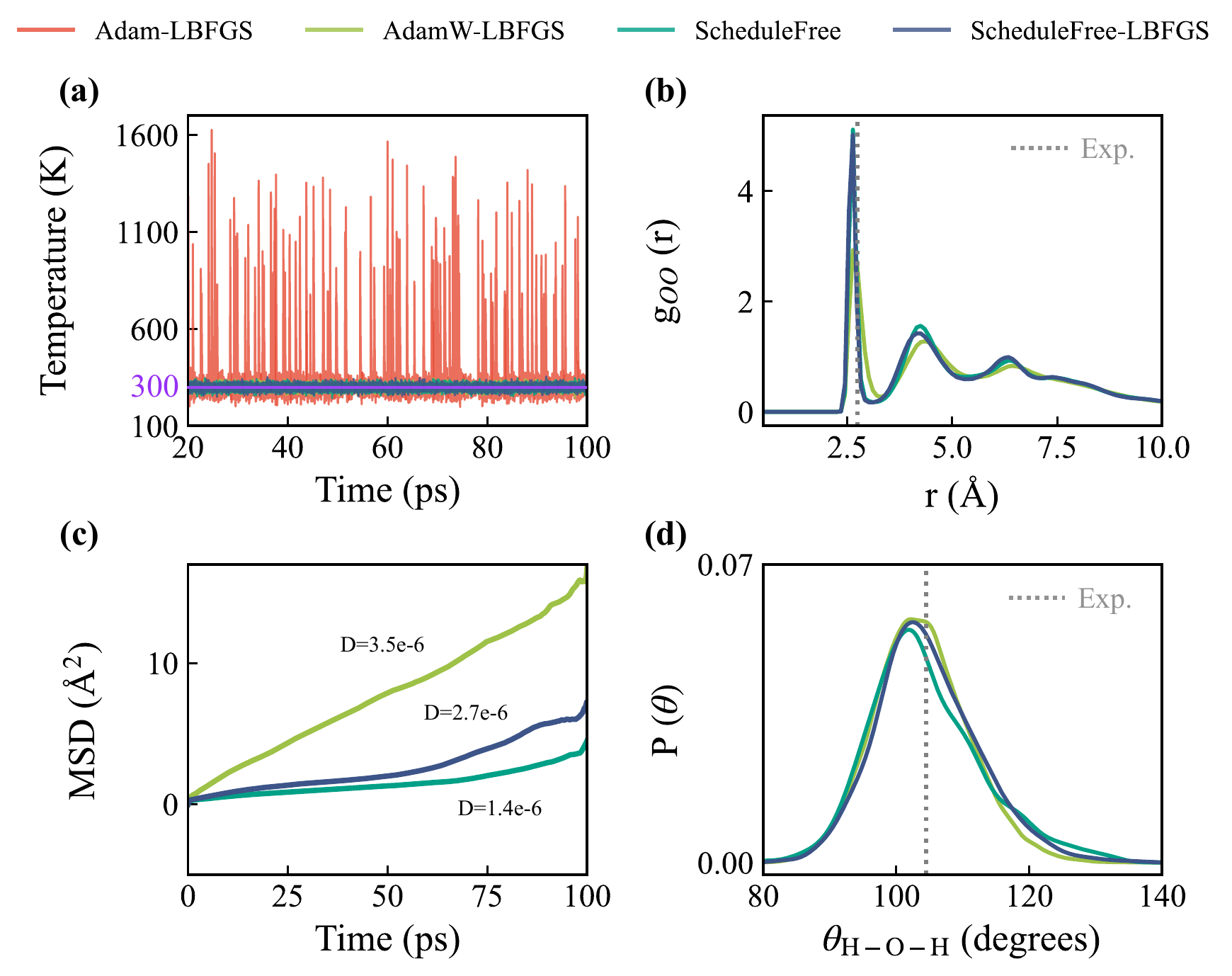}
    \caption{Comparison of MACE-MD simulations for water on graphene trained with different optimization strategies in the NVT ensemble at 300 K. {\bf (a)} Temperature evolution over a 100 ps trajectory. The model trained with Adam followed by L-BFGS (red) exhibits significant thermal fluctuations despite the presence of a thermostat. {\bf (b)} RDF of oxygen-oxygen pairs ($g_{\text{OO}}(r)$). {\bf (c)} MSD of oxygen atoms with the self-diffusion coefficient ($D$) indicated. {\bf (d)} Probability density distribution of the H-O-H bond angle ($\theta_{\text{H-O-H}}$).}
    \label{fig:gw-rdf}
\end{figure}

Figure~\ref{fig:gw-rdf}(a) demonstrates that Adam-based models, even with L-BFGS refinement, exhibit severe thermal instability that the thermostat cannot regulate. This pathology arises from an incomplete resolution of the spectral disparity between the rigid graphene lattice and the fluxional water network, which leaves residual roughness in the potential energy surface and generates impulsive forces during integration. Due to this dynamical breakdown, we exclude the Adam trajectories from the subsequent structural analysis. The AdamW and ScheduleFree optimizers, however, successfully navigate this anisotropy and maintain a stable 300 K ensemble. ScheduleFree proves particularly effective at balancing the stiff and soft manifolds, and its second-order refinement yields conservative dynamics free from the artifacts observed in the Adam-based baselines.

We further quantify the physical fidelity through structural and transport observables. The radial distribution functions in Figure~\ref{fig:gw-rdf}(b) confirm that the refined models accurately reproduce both the interfacial coordination and the bulk water structure, matching the experimental first hydration shell peak at $r \approx 2.75$ Å~\cite{soper2000radial}. Similarly, the bond angle distributions in Figure~\ref{fig:gw-rdf}(d) converge to the canonical value of $104.5^\circ$~\cite{hoy1979precise}. The transport properties in Figure~\ref{fig:gw-rdf}(c) reveal the distinct advantage of the AdamW-based refinement. While other stable configurations exhibit suppressed mobility, the model trained with AdamW and refined with L-BFGS yields a mean squared displacement slope that aligns most closely with the expected dynamics of the confined phase~\cite{ayappa2019enhancing}. This trajectory recovers a physically realistic diffusion coefficient without the artificial dynamical arrest observed in alternative schemes, establishing AdamW combined with L-BFGS as the most robust strategy for resolving the multiscale forces at heterogeneous interfaces.

\paragraph{Cost and benefit analysis.}

\begin{figure}[t]
    \centering
    \includegraphics[width=0.85\linewidth]{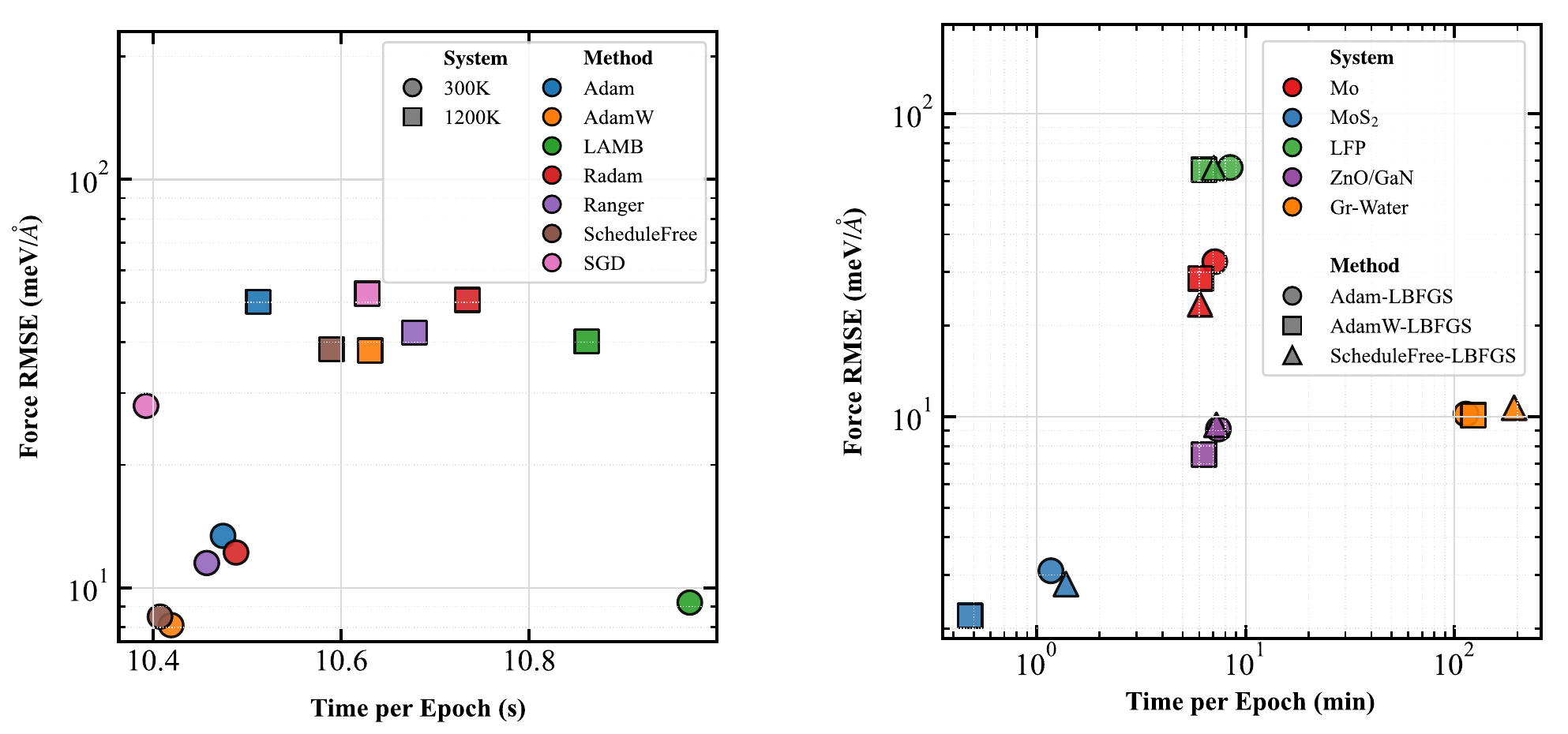}
    \caption{Pareto frontiers illustrating the trade-off between force accuracy and computational cost. \textbf{Left:} Performance of representative first-order optimizers on the 3BPA dataset, evaluated on in-distribution (300 K) and out-of-distribution (1200 K) test sets. \textbf{Right:} Impact of the second-order refinement stage across diverse material systems. The plot compares the force RMSE and training cost of the base first-order models against their L-BFGS refined counterparts, quantifying the accuracy gains relative to the computational overhead.}
    \label{fig:pareto}
\end{figure}

The integration of a second-order refinement stage requires a critical evaluation of the balance between training cost and physical fidelity. We quantify this relationship through the Pareto frontiers presented in Figure~\ref{fig:pareto}, which map the convergence of force accuracy against the wall clock time. The left panel characterizes the baseline efficiency of first-order optimizers on the 3BPA benchmark. While adaptive methods such as AdamW and ScheduleFree rapidly approach the statistical limit of the dataset, the right panel reveals the distinct economic profile of the L-BFGS refinement across different material systems.

For structurally homogeneous systems such as bulk Mo and MoS$_2$, the refinement trajectory in Figure~\ref{fig:pareto} (Right) is primarily horizontal. This indicates that the additional computational overhead incurred by line search evaluations and curvature history updates yields negligible reductions in force error. In these regimes, the potential energy surface is well conditioned, and the first-order baselines have already saturated the regression floor. Consequently, the marginal gain in accuracy does not justify the increased training cost for high throughput screening applications where static properties are the primary concern.

The scenario changes fundamentally for heterogeneous environments characterized by spectral anisotropy, such as the graphene and water interface. Although the reduction in aggregate RMSE appears moderate in the Pareto plot, this metric underestimates the true utility of the refinement. As established in Section~\ref{sec:results}, the primary contribution of L-BFGS in this regime is not only minimizing the residual error but regularizing the Hessian along soft interfacial modes that first-order methods do not resolve. This spectral correction is essential to prevent the thermal instability and unphysical artifacts observed in the baseline trajectories. Furthermore, since the refinement stage modifies only the model parameters, and not the architecture, it imposes zero additional latency during inference. We therefore recommend a stratified optimization strategy: standard first-order methods suffice for simple systems, whereas a brief second-order refinement phase is indispensable for ensuring dynamical consistency and rigorous conservation laws in complex multiscale interfaces.

\section{Discussion and Outlook}
\label{sec:discussion}

In this work, we provided a systematic and quantitative study of how optimizer choice influences the fine-tuning of atomistic foundation models. 
Through unified benchmarks covering inorganic, organic, and liquid systems, we have shown that optimization dynamics is not interchangeable, but imprints measurable biases on the resulting potential-energy surfaces. 
Even when trained under identical data and architectural conditions, different optimizers produce distinct convergence behaviors and physical fidelities, revealing that the path through parameter space is itself a determinant of model quality.

From a theoretical point of view, the preconditioning perspective developed in Section~\ref{sec:sub:preconditioning} offers a unifying interpretation of these results. The first-order optimizers can be viewed as constructing approximate inverses of the local curvature, thereby moderating the anisotropy of the loss landscape. Rectified and schedule-free variants further adjust this implicit preconditioner by adapting the spectral scale of the effective learning matrix, which leads to more uniform contraction across curvature modes. The empirical observation that these optimizers yield smoother potential-energy surfaces and more stable dynamics provides strong evidence that curvature conditioning is a key factor governing the transferability of fine-tuned interatomic potentials.

Beyond these mechanistic insights, the benchmarks identify several practical guidelines for future development. First, optimizers that combine moderate adaptivity with reliable variance estimation, such as AdamW and ScheduleFree, achieve consistent accuracy without extensive hyperparameter tuning, making them robust defaults for fine-tuning large pretrained atomistic models. 
Second, short second-order refinement stages using L-BFGS can further polish energy accuracy when required, although their computational cost limits their use to targeted applications. Finally, purely non-adaptive methods appear suboptimal for highly anisotropic and high-dimensional atomistic landscapes, emphasizing the need for curvature-aware design principles.

These findings underscore that optimization algorithms are not mere engineering details, but integral components of the modeling framework. 
Incorporating geometric information about the loss surface into the optimizer, for example through low-rank curvature updates or blockwise preconditioners respecting atomic equivariance, may represent a promising path forward. Such designs could bridge the efficiency of first-order methods with the stability of second-order schemes, aligning optimization dynamics more closely with the physical structure of the problem.
% In a broader context, the conclusions drawn here suggest that the reproducibility and generalization of machine-learned interatomic potentials depend as much on the choice of optimizer as on data coverage or model capacity. 
% Establishing standardized fine-tuning protocols and reporting optimization details transparently will therefore be essential for future benchmark comparisons and cross-study reproducibility. 
The systematic exploration of optimizer-induced effects presented in this work lays a foundation for such practices and highlights a new direction for research at the interface of machine learning optimization and atomistic modeling.

\appendix

\section{Theoretical Analysis}
\label{sec:apd:analysis}

Many first-order optimizers used for fine-tuning universal MLIPs can be interpreted as providing a diagonal, data-dependent preconditioning operator. The following result formalizes how
such diagonal scaling modifies the spectrum of the local Hessian and
clarifies the connection between the optimizer families discussed in
Section~\ref{sec:sub:preconditioning} and the numerical behaviour
reported in Section~\ref{sec:results}.

Let $H\in\mathbb{R}^{d\times d}$ be symmetric positive definite with
eigen-decomposition $H = Q\Lambda Q^{\mathsf T}$,
$\Lambda=\mathrm{diag}(\lambda_1,\dots,\lambda_d)$, where
$0<\lambda_{\min}\le\lambda_i\le\lambda_{\max}$.
Let $P$ be a positive definite diagonal preconditioner that commutes
with~$H$, so that
$P = Q\,\mathrm{diag}(p_1,\dots,p_d)\,Q^{\mathsf T}$.
Assume that for some $\alpha\in[0,1]$ and constants
$0<c_{\min}\le c_{\max}<\infty$,
\begin{equation}
c_{\min}\,\lambda_i^{-\alpha}
\le
p_i
\le
c_{\max}\,\lambda_i^{-\alpha},
\qquad i=1,\dots,d .
\end{equation}
Then the condition number of the preconditioned operator satisfies
\begin{equation}
\label{eq:S1-bounds-updated}
\frac{c_{\min}}{c_{\max}}\,
\kappa(H)^{\,1-\alpha}
\le
\kappa(PH)
\le
\frac{c_{\max}}{c_{\min}}\,
\kappa(H)^{\,1-\alpha},
\end{equation}
and, in the idealized case $P = c\,H^{-\alpha}$,
\begin{equation}
\label{eq:S1-exact-updated}
\kappa(PH)=\kappa(H)^{\,1-\alpha}.
\end{equation}

Special cases are recovered by taking $\alpha=0$ (unpreconditioned gradient
descent), $\alpha=1$ (Newton scaling), or intermediate $\alpha\in(0,1)$,
which corresponds to varying degrees of diagonal spectral flattening.

% \emph{Proof}.
% Because $P$ and $H$ are simultaneously diagonalizable, the eigenvalues of
% $PH$ are $\mu_i = p_i\lambda_i$. The assumed bounds imply
% $c_{\min}\lambda_i^{1-\alpha} \le \mu_i \le c_{\max}\lambda_i^{1-\alpha}$.
% Taking maxima and minima over~$i$ yields the two-sided bound
% \eqref{eq:S1-bounds-updated}.  
% The identity \eqref{eq:S1-exact-updated} follows directly when
% $p_i=c\,\lambda_i^{-\alpha}$.  \hfill$\square$

% \textit{Remarks.}
The parameter $\alpha$ provides a compact way to describe the effective
spectral action of several optimizers. In quadratic neighbourhoods, the
variance estimate in Adam-type methods satisfies
$v_{t,i}\approx \mathbb{E}[g_{t,i}^2]\propto\lambda_i^2$, so the scaling
$(\sqrt{v_{t,i}}+\varepsilon)^{-1}$ behaves like $\lambda_i^{-1}$,
corresponding to $\alpha\approx 1$.  
RAdam interpolates between $\alpha\approx 0$ and $\alpha\approx 1$
as the variance estimate stabilizes.  
Layerwise rescaling in LAMB preserves the elementwise structure while
normalizing the update magnitude at the block level, leaving the effective
$\alpha$ essentially unchanged.  
ScheduleFree introduces a slowly varying scalar multiplier that does
not alter the exponent~$\alpha$ but adjusts the overall contraction radius.  
Finally, L-BFGS recovers $H^{-1}$ on the Krylov subspace generated by recent
gradients, yielding exact $\alpha=1$ on that subspace.

These observations clarify the empirical behaviour reported in
Section~\ref{sec:results}.  
Universal MLIPs exhibit large and heterogeneous condition numbers due to
the coupling of equivariant layers and energy–force supervision.  
Improving the effective exponent $\alpha$ therefore leads to substantial
reductions in the effective condition number $\kappa(PH)$ and, consequently,
to more stable and efficientfine-tuning.  
The combination of an adaptive first-order method (which provides
$\alpha$ close to~1 diagonally) followed by a brief L-BFGS refinement
(which enforces $\alpha=1$ on a low-dimensional subspace) is consistent
with the observed improvements in the terminal phase of training.

\section{L-BFGS refinement procedure}
\label{sec:apd:lbfgs}

This appendix details the second-order refinement stage that we apply after
first-order fine-tuning. Given parameters \(\theta_T\) obtained from a chosen
optimizer, we run a L-BFGS procedure on the same
objective \(f(\theta) = \mathcal{L}(\theta)\). Each iteration performs a backtracking line-search that
enforces Armijo–Wolfe conditions~\cite{jin2024non}, updates the curvature history with Powell
damping to preserve positive definiteness, and monitors validation force RMSE
for early stopping. The output of this stage is the checkpoint with the best
validation force RMSE observed during refinement.

\begin{algorithm}[h]
\caption{L-BFGS refinement after first-order fine-tuning}
\label{alg:lbfgs_refine}
\begin{algorithmic}[1]
\State \textbf{Input:}
initial parameters \(\theta_0 = \theta_T\) (from first-order run),
objective \(f(\theta)\),
memory size \(m\),
maximum refinement steps \(K\)
\State Compute \(g_0 = \nabla f(\theta_0)\); initialize curvature history \(\mathcal{H} \gets \varnothing\)
\State Initialize best checkpoint \(\theta_{\text{best}} \gets \theta_0\) and corresponding validation force RMSE

\For{\(k = 0,1,\dots,K-1\)}
    \State \textbf{// search direction via two-loop recursion}
    \State \(p_k \gets \textsc{TwoLoopRecursion}(g_k, \mathcal{H})\) \Comment{standard L-BFGS with at most \(m\) pairs}
    
    \State \textbf{// backtracking line-search}
    \State Choose step size \(\alpha_k\) by backtracking along \(p_k\)
           until Armijo–Wolfe conditions are satisfied
    \State \(\theta_{k+1} \gets \theta_k + \alpha_k p_k\)
    \State \(g_{k+1} \gets \nabla f(\theta_{k+1})\)
    
    \State \textbf{// update curvature pairs with Powell damping}
    \State \(s_k \gets \theta_{k+1} - \theta_k\), \quad \(y_k \gets g_{k+1} - g_k\)
    \If{\(y_k^{\mathsf T} s_k \le 0\)}
        \State apply Powell damping to \((s_k, y_k)\) to enforce \(y_k^{\mathsf T} s_k > 0\)
    \EndIf
    \State Append \((s_k, y_k)\) to \(\mathcal{H}\) and discard the oldest pair if \(|\mathcal{H}| > m\)
    
    \State \textbf{// early stopping based on validation error}
    \State Evaluate validation force RMSE at \(\theta_{k+1}\)
    \If{validation force RMSE improves}
        \State update \(\theta_{\text{best}}\) and reset patience counter
    \Else
        \State increase patience counter; \textbf{break} if patience exceeds a prescribed threshold
    \EndIf
    
    \State \textbf{// additional safeguard}
    \If{\(\|p_k\|\) is below a prescribed tolerance}
        \State \textbf{break}
    \EndIf
\EndFor

\State \textbf{Output:} refined parameters \(\theta_{\text{best}}\)
\end{algorithmic}
\end{algorithm}

\section{Ablation Study}
\label{sec:apd:ablation}

To assess the sensitivity of the different optimizers to the choice of learning rate, we conduct an ablation study on the 3BPA dataset. The models are trained on configurations at \(300\,\mathrm{K}\) and evaluated on test sets at \(300\,\mathrm{K}\) and \(1200\,\mathrm{K}\). Table~\ref{tab:ablation} reports the RMSE for energies and forces for learning rates of \(5\times 10^{-3}\), \(1\times 10^{-3}\), and \(5\times 10^{-4}\).

Across all temperatures and prediction targets, AdamW and ScheduleFree consistently achieve the lowest errors and remain robust under variation of the learning rate, which is consistent with the observations reported in the main text. AdamW attains the best or second best RMSE in every setting and typically reaches its minimum error at the largest learning rate, while ScheduleFree closely matches this performance and exhibits similarly stable behavior. In contrast, SGD produces substantially larger errors and degrades significantly when the learning rate increases, and the remaining adaptive methods, namely Adam, RAdam, LAMB, and Ranger, perform between these two extremes. These results indicate that the superiority of AdamW and ScheduleFree does not arise from a particular tuning of the learning rate, but instead reflects more favorable optimization dynamics and more effective preconditioning of the loss landscape.

\begin{table*}[h]
\caption{RMSE of energy (E, meV/atom) and force (F, meV/\AA) on the 3BPA dataset with different learning rates. 
The training set is collected at 300 K. The best two results of each conditions are in bold.}
\label{tab:ablation}
\begin{center}
\begin{scriptsize}
\renewcommand{\arraystretch}{1.4} % 调大行距
\begin{tabular}{lcccccccc}
\toprule
{\bf Condition} & {\bf Value} & {\bf SGD} & {\bf Adam} & {\bf AdamW} & {\bf RAdam} & {\bf LAMB} & {\bf Ranger} & {\bf ScheduleFree} \\
\midrule
\multirow{3}{*}{300K, E} 
 & $5\times 10^{-3}$ & 0.6 & 0.4 & {\bf 0.1} & 0.5 & 0.4 & 0.2 & {\bf 0.1} \\
 & $1\times 10^{-3}$ & 0.7 & {\bf 0.2} & {\bf 0.2} & 0.6 & {\bf 0.2} & 0.3 & {\bf 0.2} \\
 & $5\times 10^{-4}$ & 1.2 & 0.5 & {\bf 0.2} & 0.4 & {\bf 0.2} & 0.3 & {\bf 0.2} \\
\hline
\multirow{3}{*}{300K, F} 
 & $5\times 10^{-3}$ & 20.5  & 14.5 & {\bf 7.4} & 12.8 & 8.8 & 9.9 & {\bf 8.1} \\
 & $1\times 10^{-3}$ & 27.9  & 11.2 & {\bf 8.1} & 12.2 & 9.2 & 11.5 & {\bf 8.5} \\
 & $5\times 10^{-4}$ & 36.1  & 12.1 & {\bf 8.6} & 11.9 & 9.8 & 12.2 & {\bf 9.0} \\
\hline
\multirow{3}{*}{1200K, E} 
 & $5\times 10^{-3}$ & 1.2 & 1.1 & {\bf 0.6} & 0.9 & 0.8 & 0.8 & {\bf 0.7} \\
 & $1\times 10^{-3}$ & 1.4 & 0.9 & {\bf 0.7} & 0.9 & 0.9 & {\bf 0.7} & {\bf 0.6} \\
 & $5\times 10^{-4}$ & 2.1 & 0.9 & {\bf 0.5} & 0.9 & 0.9 & 0.8 & {\bf 0.7} \\
 \hline
\multirow{3}{*}{1200K, F} 
 & $5\times 10^{-3}$ & 47.7 & 55.9 & {\bf 38.3} & 49.5 & 40.0 & 40.9 & {\bf 39.4} \\
 & $1\times 10^{-3}$ & 52.5 & 50.0 & {\bf 38.2} & 51.5 & 40.4 & 42.3 & {\bf 38.4} \\
 & $5\times 10^{-4}$ & 57.1 & 51.0 & {\bf 38.5} & 50.3 & 49.5 & 42.5 & {\bf 39.0} \\
\bottomrule
\end{tabular}
\end{scriptsize}
\end{center}
\end{table*}

\vspace{1cm}

\bibliographystyle{plain} 
\bibliography{bib.bib}

@article{chen2022universal,
  title={A universal graph deep learning interatomic potential for the periodic table},
  author={Chen, Chi and Ong, Shyue Ping},
  journal={Nat. Comput. Sci.},
  volume={2},
  number={11},
  pages={718--728},
  year={2022},
  publisher={Nature Publishing Group US New York}
}

@article{musil2021physics,
  title={Physics-inspired structural representations for molecules and materials},
  author={Musil, Felix and Grisafi, Andrea and Bart{\'o}k, Albert P and Ortner, Christoph and Cs{\'a}nyi, G{\'a}bor and Ceriotti, Michele},
  journal={Chem. Rev.},
  volume={121},
  number={16},
  pages={9759--9815},
  year={2021},
  publisher={ACS Publications}
}

@article{jacobs2025practical,
  title={A practical guide to machine learning interatomic potentials--Status and future},
  author={Jacobs, Ryan and Morgan, Dane and Attarian, Siamak and Meng, Jun and Shen, Chen and Wu, Zhenghao and Xie, Clare Yijia and Yang, Julia H and Artrith, Nongnuch and Blaiszik, Ben and others},
  journal={Curr. Opin. Solid State Mater. Sci.},
  volume={35},
  pages={101214},
  year={2025},
  publisher={Elsevier}
}

@article{botu2017machine,
  title={Machine learning force fields: construction, validation, and outlook},
  author={Botu, Venkatesh and Batra, Rohit and Chapman, James and Ramprasad, Rampi},
  journal={J. Phys. Chem. C},
  volume={121},
  number={1},
  pages={511--522},
  year={2017},
  publisher={ACS Publications}
}

@article{barroso2024open,
  title={Open materials 2024 (omat24) inorganic materials dataset and models},
  author={Barroso-Luque, Luis and Shuaibi, Muhammed and Fu, Xiang and Wood, Brandon M and Dzamba, Misko and Gao, Meng and Rizvi, Ammar and Zitnick, C Lawrence and Ulissi, Zachary W},
  journal={arXiv preprint arXiv:2410.12771},
  year={2024}
}

@article{poltavsky2021machine,
  title={Machine learning force fields: Recent advances and remaining challenges},
  author={Poltavsky, Igor and Tkatchenko, Alexandre},
  journal={J. Phys. Chem. Lett.},
  volume={12},
  number={28},
  pages={6551--6564},
  year={2021},
  publisher={ACS Publications}
}

@article{unke2021machine,
  title={Machine learning force fields},
  author={Unke, Oliver T and Chmiela, Stefan and Sauceda, Huziel E and Gastegger, Michael and Poltavsky, Igor and Schutt, Kristof T and Tkatchenko, Alexandre and Muller, Klaus-Robert},
  journal={Chem. Rev.},
  volume={121},
  number={16},
  pages={10142--10186},
  year={2021},
  publisher={ACS Publications}
}

@article{thompson2022lammps,
  title={{LAMMPS}-a flexible simulation tool for particle-based materials modeling at the atomic, meso, and continuum scales},
  author={Thompson, Aidan P and Aktulga, H Metin and Berger, Richard and Bolintineanu, Dan S and Brown, W Michael and Crozier, Paul S and In't Veld, Pieter J and Kohlmeyer, Axel and Moore, Stan G and Nguyen, Trung Dac and others},
  journal={Comput. Phys. Commun.},
  volume={271},
  pages={108171},
  year={2022},
  publisher={Elsevier}
}

@article{xie2023ultra,
  title={Ultra-fast interpretable machine-learning potentials},
  author={Xie, Stephen R and Rupp, Matthias and Hennig, Richard G},
  journal={npj Comput. Mater.},
  volume={9},
  number={1},
  pages={162},
  year={2023},
  publisher={Nature Publishing Group UK London}
}

@article{thompson2015spectral,
  title={Spectral neighbor analysis method for automated generation of quantum-accurate interatomic potentials},
  author={Thompson, Aidan P and Swiler, Laura P and Trott, Christian R and Foiles, Stephen M and Tucker, Garritt J},
  journal={J. Comput. Phys.},
  volume={285},
  pages={316--330},
  year={2015},
  publisher={Elsevier}
}

@article{bochkarev2024graph,
  title={Graph atomic cluster expansion for semilocal interactions beyond equivariant message passing},
  author={Bochkarev, Anton and Lysogorskiy, Yury and Drautz, Ralf},
  journal={Phys. Rev. X},
  volume={14},
  number={2},
  pages={021036},
  year={2024},
  publisher={APS}
}

@article{smith2017ani,
  title={{ANI}-1: an extensible neural network potential with {DFT} accuracy at force field computational cost},
  author={Smith, Justin S and Isayev, Olexandr and Roitberg, Adrian E},
  journal={Chem. Sci.},
  volume={8},
  number={4},
  pages={3192--3203},
  year={2017},
  publisher={Royal Society of Chemistry}
}

@article{musaelian2023learning,
  title={Learning local equivariant representations for large-scale atomistic dynamics},
  author={Musaelian, Albert and Batzner, Simon and Johansson, Anders and Sun, Lixin and Owen, Cameron J and Kornbluth, Mordechai and Kozinsky, Boris},
  journal={Nat. Commun.},
  volume={14},
  number={1},
  pages={579},
  year={2023},
  publisher={Nature Publishing Group UK London}
}

@article{cheng2024cartesian,
  title={Cartesian atomic cluster expansion for machine learning interatomic potentials},
  author={Cheng, Bingqing},
  journal={npj Comput. Mater.},
  volume={10},
  number={1},
  pages={157},
  year={2024},
  publisher={Nature Publishing Group UK London}
}

@article{batzner20223,
  title={E (3)-equivariant graph neural networks for data-efficient and accurate interatomic potentials},
  author={Batzner, Simon and Musaelian, Albert and Sun, Lixin and Geiger, Mario and Mailoa, Jonathan P and Kornbluth, Mordechai and Molinari, Nicola and Smidt, Tess E and Kozinsky, Boris},
  journal={Nat. Commun.},
  volume={13},
  number={1},
  pages={2453},
  year={2022},
  publisher={Nature Publishing Group UK London}
}

@article{wang2018deepmd,
  title={Dee{PMD}-kit: A deep learning package for many-body potential energy representation and molecular dynamics},
  author={Wang, Han and Zhang, Linfeng and Han, Jiequn and others},
  journal={Comput. Phys. Commun.},
  volume={228},
  pages={178--184},
  year={2018},
  publisher={Elsevier}
}

@article{choudhary2023unified,
  title={Unified graph neural network force-field for the periodic table: solid state applications},
  author={Choudhary, Kamal and DeCost, Brian and Major, Lily and Butler, Keith and Thiyagalingam, Jeyan and Tavazza, Francesca},
  journal={Digit. Discov.},
  volume={2},
  number={2},
  pages={346--355},
  year={2023},
  publisher={Royal Society of Chemistry}
}

@article{bartok2018machine,
  title={Machine learning a general-purpose interatomic potential for silicon},
  author={Bart{\'o}k, Albert P and Kermode, James and Bernstein, Noam and Cs{\'a}nyi, G{\'a}bor},
  journal={Phys. Rev. X},
  volume={8},
  number={4},
  pages={041048},
  year={2018},
  publisher={APS}
}

@article{behler2007generalized,
  title={Generalized neural-network representation of high-dimensional potential-energy surfaces},
  author={Behler, J{\"o}rg and Parrinello, Michele},
  journal={Phys. Rev. Lett.},
  volume={98},
  number={14},
  pages={146401},
  year={2007},
  publisher={APS}
}

@article{bartok2010gaussian,
  title={Gaussian approximation potentials: The accuracy of quantum mechanics, without the electrons},
  author={Bart{\'o}k, Albert P and Payne, Mike C and Kondor, Risi and Cs{\'a}nyi, G{\'a}bor},
  journal={Phys. Rev. Lett.},
  volume={104},
  number={13},
  pages={136403},
  year={2010},
  publisher={APS}
}

@article{batatia2023foundation,
  title={A foundation model for atomistic materials chemistry},
  author={Batatia, Ilyes and Benner, Philipp and Chiang, Yuan and Elena, Alin M and Kov{\'a}cs, D{\'a}vid P and Riebesell, Janosh and Advincula, Xavier R and Asta, Mark and Baldwin, William J and Bernstein, Noam and others},
  journal={arXiv preprint arXiv:2401.00096},
  year={2023}
}

@article{batatia2022mace,
  title={M{ACE}: Higher order equivariant message passing neural networks for fast and accurate force fields},
  author={Batatia, Ilyes and Kovacs, David P and Simm, Gregor and Ortner, Christoph and Cs{\'a}nyi, G{\'a}bor},
  journal={Adv. Neural Inf. Process. Syst.},
  volume={35},
  year={2022}
}

@article{bowman2022md17,
  title={The {MD17} datasets from the perspective of datasets for gas-phase “small” molecule potentials},
  author={Bowman, Joel M and Qu, Chen and Conte, Riccardo and Nandi, Apurba and Houston, Paul L and Yu, Qi},
  journal={J. Chem. Phys.},
  volume={156},
  number={24},
  year={2022},
  publisher={AIP Publishing}
}

@article{chanussot2021open,
  title={Open catalyst 2020 {(OC20)} dataset and community challenges},
  author={Chanussot, Lowik and Das, Abhishek and Goyal, Siddharth and Lavril, Thibaut and Shuaibi, Muhammed and Riviere, Morgane and Tran, Kevin and Heras-Domingo, Javier and Ho, Caleb and Hu, Weihua and others},
  journal={ACS Catal.},
  volume={11},
  number={10},
  pages={6059--6072},
  year={2021},
  publisher={ACS Publications}
}

@article{pyzer2025foundation,
  title={Foundation models for materials discovery--current state and future directions},
  author={Pyzer-Knapp, Edward O and Manica, Matteo and Staar, Peter and Morin, Lucas and Ruch, Patrick and Laino, Teodoro and Smith, John R and Curioni, Alessandro},
  journal={npj Comput. Mater.},
  volume={11},
  number={1},
  pages={61},
  year={2025},
  publisher={Nature Publishing Group UK London}
}

@article{radova2025fine,
  title={Fine-tuning foundation models of materials interatomic potentials with frozen transfer learning},
  author={Radova, Mariia and Stark, Wojciech G and Allen, Connor S and Maurer, Reinhard J and Bart{\'o}k, Albert P},
  journal={npj Computational Materials},
  volume={11},
  number={1},
  pages={237},
  year={2025},
  publisher={Nature Publishing Group UK London}
}

@article{lee2025accelerating,
  title={Accelerating high-throughput phonon calculations via machine learning universal potentials},
  author={Lee, Huiju and Hegde, Vinay I and Wolverton, Chris and Xia, Yi},
  journal={Mater. Today Phys.},
  volume={53},
  pages={101688},
  year={2025},
  publisher={Elsevier}
}

@article{du2025universal,
  title={Universal Machine Learning Interatomic Potentials are Ready for Solid Ion Conductors},
  author={Du, Hongwei and Hui, Jian and Zhang, Lanting and Wang, Hong},
  journal={arXiv preprint arXiv:2502.09970},
  year={2025}
}

@article{casillas2024evaluating,
  title={Evaluating and improving the predictive accuracy of mixing enthalpies and volumes in disordered alloys from universal pretrained machine learning potentials},
  author={Casillas-Trujillo, Luis and Parackal, Abhijith S and Armiento, Rickard and Alling, Bj{\"o}rn},
  journal={Phys. Rev. Mater.},
  volume={8},
  number={11},
  pages={113803},
  year={2024},
  publisher={APS}
}

@article{niblett2024transferability,
  title={Transferability of datasets between Machine-Learning Interaction Potentials},
  author={Niblett, Samuel P and Kourtis, Panagiotis and Magd{\u{a}}u, Ioan-Bogdan and Grey, Clare P and Cs{\'a}nyi, G{\'a}bor},
  journal={arXiv preprint arXiv:2409.05590},
  year={2024}
}

@article{shuang2025universal,
  title={Universal machine learning interatomic potentials poised to supplant {DFT} in modeling general defects in metals and random alloys},
  author={Shuang, Fei and Wei, Zixiong and Liu, Kai and Gao, Wei and Dey, Poulumi},
  journal={arXiv preprint arXiv:2502.03578},
  year={2025}
}

@article{yu2024systematic,
  title={Systematic assessment of various universal machine-learning interatomic potentials},
  author={Yu, Haochen and Giantomassi, Matteo and Materzanini, Giuliana and Wang, Junjie and Rignanese, Gian-Marco},
  journal={Mater. Genome Eng. Adv.},
  volume={2},
  number={3},
  pages={e58},
  year={2024},
  publisher={Wiley Online Library}
}

@article{yang2024mattersim,
  title={Mattersim: A deep learning atomistic model across elements, temperatures and pressures},
  author={Yang, Han and Hu, Chenxi and Zhou, Yichi and Liu, Xixian and Shi, Yu and Li, Jielan and Li, Guanzhi and Chen, Zekun and Chen, Shuizhou and Zeni, Claudio and others},
  journal={arXiv preprint arXiv:2405.04967},
  year={2024}
}

@article{zhang2024dpa,
  title={{DPA}-2: a large atomic model as a multi-task learner},
  author={Zhang, Duo and Liu, Xinzijian and Zhang, Xiangyu and Zhang, Chengqian and Cai, Chun and Bi, Hangrui and Du, Yiming and Qin, Xuejian and Peng, Anyang and Huang, Jiameng and others},
  journal={npj Comput. Mater.},
  volume={10},
  number={1},
  pages={293},
  year={2024},
  publisher={Nature Publishing Group UK London}
}

@article{deng2023chgnet,
  title={CHGNet as a pretrained universal neural network potential for charge-informed atomistic modelling},
  author={Deng, Bowen and Zhong, Peichen and Jun, KyuJung and Riebesell, Janosh and Han, Kevin and Bartel, Christopher J and Ceder, Gerbrand},
  journal={Nat. Mach. Intell.},
  volume={5},
  number={9},
  pages={1031--1041},
  year={2023},
  publisher={Nature Publishing Group UK London}
}

@Article{DrautzACE,
  title = {Atomic cluster expansion for accurate and transferable interatomic potentials},
  author = {Drautz, Ralf},
  journal = {Phys. Rev. B},
  volume = {99},
  issue = {1},
  pages = {014104},
  numpages = {15},
  year = {2019},
  month = {Jan},
  publisher = {American Physical Society}
}

@article{naghdi2024neural,
  title={Neural network interatomic potentials for open surface nano-mechanics applications},
  author={Naghdi, A. D. and Pellegrini, F. and K{\"u}{\c{c}}{\"u}kbenli, E. and others},
  journal={Acta Mater.},
  volume={277},
  pages={120200},
  year={2024},
  publisher={Elsevier}
}

@Article{ACECompleteness,
  author    = {Dusson, Genevi{\`e}ve and Bachmayr, Markus and Cs{\'a}nyi, G{\'a}bor and Drautz, Ralf and Etter, Simon and van der Oord, Cas and Ortner, Christoph},
  journal   = {J. Comput. Phys.},
  title     = {Atomic cluster expansion: Completeness, efficiency and stability},
  year      = {2022},
  pages     = {110946},
  volume    = {454},
  fjournal  = {J. Comput. Phys.},
  publisher = {Elsevier},
}

@article{kingma2014adam,
  title={A method for stochastic optimization},
  author={Adam, Kingma DP Ba J and others},
  journal={arXiv preprint arXiv:1412.6980},
  volume={1412},
  number={6},
  year={2014}
}

@article{merchant2023scaling,
  title={Scaling deep learning for materials discovery},
  author={Merchant, Amil and Batzner, Simon and Schoenholz, Samuel S and Aykol, Muratahan and Cheon, Gowoon and Cubuk, Ekin Dogus},
  journal={Nature},
  volume={624},
  number={7990},
  pages={80--85},
  year={2023},
  publisher={Nature Publishing Group UK London}
}

@article{perdew1996generalized,
  title={Generalized gradient approximation made simple},
  author={Perdew, John P and Burke, Kieron and Ernzerhof, Matthias},
  journal={Phys. Rev. Lett.},
  volume={77},
  number={18},
  pages={3865},
  year={1996},
  publisher={APS}
}

@article{deng2024overcoming,
  title={Systematic softening in universal machine learning interatomic potentials},
  author={Deng, Bowen and Choi, Yunyeong and Zhong, Peichen and Riebesell, Janosh and Anand, Shashwat and Li, Zhuohan and Jun, KyuJung and Persson, Kristin A and Ceder, Gerbrand},
  journal={npj Comput. Mater.},
  volume={11},
  number={1},
  pages={1--9},
  year={2025},
  publisher={Nature Publishing Group}
}

@article{focassio2024performance,
  title={Performance assessment of universal machine learning interatomic potentials: Challenges and directions for materials’ surfaces},
  author={Focassio, Bruno and M. Freitas, Luis Paulo and Schleder, Gabriel R},
  journal={ACS Appl. Mater. Interfaces},
  year={2024},
  volume   = {17},
  pages = {13111--12121},
  publisher={ACS Publications}
}

@article{schutt2017schnet,
  title={Schnet: A continuous-filter convolutional neural network for modeling quantum interactions},
  author={Sch{\"u}tt, Kristof and Kindermans, Pieter-Jan and Sauceda Felix, Huziel Enoc and Chmiela, Stefan and Tkatchenko, Alexandre and Muller, Klaus-Robert},
  journal={Adv. Neural Inf. Process. Syst.},
  volume={30},
  pages={992--1002},
  year={2017}
}

@article{shapeev2016moment,
  title={Moment tensor potentials: A class of systematically improvable interatomic potentials},
  author={Shapeev, Alexander V},
  journal={Multiscale Model. Simul.},
  volume={14},
  number={3},
  pages={1153--1173},
  year={2016},
  publisher={SIAM}
}

@article{zhang2022dpa,
  title={Dpa-1: Pretraining of attention-based deep potential model for molecular simulation},
  author={Zhang, Duo and Bi, Hangrui and Dai, Fu-Zhi and Jiang, Wanrun and Zhang, Linfeng and Wang, Han},
  journal={arXiv preprint arXiv:2208.08236},
  year={2022}
}

@article{qu2024importance,
  title={The importance of being scalable: Improving the speed and accuracy of neural network interatomic potentials across chemical domains},
  author={Qu, Eric and Krishnapriyan, Aditi},
  journal={Advances in Neural Information Processing Systems},
  volume={37},
  pages={139030--139053},
  year={2024}
}

@article{bowman2022spectral,
  title={Spectral bias outside the training set for deep networks in the kernel regime},
  author={Bowman, Benjamin and Montufar, Guido F},
  journal={Advances in Neural Information Processing Systems},
  volume={35},
  pages={30362--30377},
  year={2022}
}

@article{qi2024robust,
  title        = {Robust training of machine learning interatomic potentials with dimensionality reduction and stratified sampling},
  author       = {Qi, Ji and Ko, Tsz Wai and Wood, Brandon C. and Pham, Tuan Anh and Ong, Shyue Ping},
  journal      = {npj Computational Materials},
  year         = {2024},
  volume       = {10},
  number       = {1},
  pages        = {55},
  doi          = {10.1038/s41524-024-01227-4},
  publisher    = {Nature Publishing Group}
}

@article{anstine2023machine,
  title        = {Machine learning interatomic potentials and long-range physics},
  author       = {Anstine, D. M. and Isayev, Olexandr},
  journal      = {The Journal of Physical Chemistry A},
  year         = {2023},
  volume       = {127},
  number       = {11},
  pages        = {2417--2431},
  doi          = {10.1021/acs.jpca.2c06778},
  publisher    = {American Chemical Society}
}

@article{eckhoff2023lifelong,
  title        = {Lifelong Machine Learning Potentials},
  author       = {Eckhoff, Marco and Reiher, Markus},
  journal      = {Journal of Chemical Theory and Computation},
  year         = {2023},
  volume       = {19},
  number       = {12},
  pages        = {4001--4019},
  doi          = {10.1021/acs.jctc.3c00137},
  publisher    = {American Chemical Society}
}

@article{hu2023rlekf,
  title        = {{RLEKF}: An Optimizer for Deep Potential with \textit{Ab Initio} Accuracy},
  author       = {Hu, Siyu and Zhang, Wentao and Sha, Qiuchen and Pan, Feng and Wang, Lin-Wang and Jia, Weile and Tan, Guangming and Zhao, Tong},
  journal      = {Computer Physics Communications},
  year         = {2024},
  volume       = {298},
  pages        = {109112},
  doi          = {10.1016/j.cpc.2023.109112},
  note         = {Preprint available as arXiv:2212.06989},
  publisher    = {Elsevier}
}

@inproceedings{llugsi2021comparison,
  title        = {Comparison between {Adam}, {AdaMax} and {AdamW} optimizers to implement a weather forecast based on neural networks for the Andean city of Quito},
  author       = {Llugsi, Ricardo and El Yacoubi, Samia and Fontaine, Adrien and Lupera, Pa{\'u}l},
  booktitle    = {2021 IEEE Fifth Ecuador Technical Chapters Meeting (ETCM)},
  year         = {2021},
  pages        = {1--6},
  doi          = {10.1109/ETCM53931.2021.9626329},
  publisher    = {IEEE}
}

@article{liu2025fine,
  title={Fine-Tuning Universal Machine-Learned Interatomic Potentials: A Tutorial on Methods and Applications},
  author={Liu, Xiaoqing and Zeng, Kehan and Luo, Zedong and Wang, Yangshuai and Zhao, Teng and Xu, Zhenli},
  journal={arXiv preprint arXiv:2506.21935},
  year={2025}
}

@article{dozat2016incorporating,
  title={Incorporating nesterov momentum into adam},
  author={Dozat, Timothy},
  year={2016}
}

@article{defazio2024road,
  title={The road less scheduled},
  author={Defazio, Aaron and Yang, Xingyu and Mehta, Harsh and Mishchenko, Konstantin and Khaled, Ahmed and Cutkosky, Ashok},
  journal={Advances in Neural Information Processing Systems},
  volume={37},
  pages={9974--10007},
  year={2024}
}

@inproceedings{liu2021learning,
  title={Learning by turning: Neural architecture aware optimisation},
  author={Liu, Yang and Bernstein, Jeremy and Meister, Markus and Yue, Yisong},
  booktitle={International conference on machine learning},
  pages={6748--6758},
  year={2021},
  organization={PMLR}
}

@article{loshchilov2017decoupled,
  title={Decoupled weight decay regularization},
  author={Loshchilov, Ilya and Hutter, Frank},
  journal={arXiv preprint arXiv:1711.05101},
  year={2017}
}

@misc{macefoundations-github,
  title        = {MACE foundations},
  author       = {{ACEsuit developers}},
  howpublished = {\url{https://github.com/ACEsuit/mace-foundations}},
  note         = {Accessed: 2025-11-29}
}

@article{batatia2025cross,
  title={Cross Learning between Electronic Structure Theories for Unifying Molecular, Surface, and Inorganic Crystal Foundation Force Fields},
  author={Batatia, Ilyes and Lin, Chen and Hart, Joseph and Kasoar, Elliott and Elena, Alin M and Norwood, Sam Walton and Wolf, Thomas and Csanyi, Gabor},
  journal={arXiv preprint arXiv:2510.25380},
  year={2025}
}

@inproceedings{choi2019empirical,
  title     = {On Empirical Comparisons of Optimizers for Deep Learning},
  author    = {Choi, Dami and Shallue, Christopher J. and Nado, Zachary and Lee, Jaehoon and Maddison, Chris J. and Dahl, George E.},
  booktitle = {Proceedings of the 8th International Conference on Learning Representations},
  year      = {2020},
  note      = {arXiv:1910.05446},
}

@article{liu2019variance,
  title={On the variance of the adaptive learning rate and beyond},
  author={Liu, Liyuan and Jiang, Haoming and He, Pengcheng and Chen, Weizhu and Liu, Xiaodong and Gao, Jianfeng and Han, Jiawei},
  journal={arXiv preprint arXiv:1908.03265},
  year={2019}
}

@article{loshchilov2016sgdr,
  title={Sgdr: Stochastic gradient descent with warm restarts},
  author={Loshchilov, Ilya and Hutter, Frank},
  journal={arXiv preprint arXiv:1608.03983},
  year={2016}
}

@article{you2019large,
  title={Large batch optimization for deep learning: Training bert in 76 minutes},
  author={You, Yang and Li, Jing and Reddi, Sashank and Hseu, Jonathan and Kumar, Sanjiv and Bhojanapalli, Srinadh and Song, Xiaodan and Demmel, James and Keutzer, Kurt and Hsieh, Cho-Jui},
  journal={arXiv preprint arXiv:1904.00962},
  year={2019}
}

@article{tong2022calibrating,
  title={Calibrating the adaptive learning rate to improve convergence of ADAM},
  author={Tong, Qianqian and Liang, Guannan and Bi, Jinbo},
  journal={Neurocomputing},
  volume={481},
  pages={333--356},
  year={2022},
  publisher={Elsevier}
}

@article{zhang2019lookahead,
  title={Lookahead optimizer: k steps forward, 1 step back},
  author={Zhang, Michael and Lucas, James and Ba, Jimmy and Hinton, Geoffrey E},
  journal={Advances in neural information processing systems},
  volume={32},
  year={2019}
}

@article{liu2025study,
  title={A Study on the Fine-Tuning Performance of Universal Machine-Learned Interatomic Potentials (U-MLIPs)},
  author={Liu, Xiaoqing and Zeng, Kehan and Wang, Yangshuai and Zhao, Teng},
  journal={arXiv preprint arXiv:2506.07401},
  year={2025}
}

@article{wu2017towards,
  title={Towards understanding generalization of deep learning: Perspective of loss landscapes},
  author={Wu, Lei and Zhu, Zhanxing and others},
  journal={arXiv preprint arXiv:1706.10239},
  year={2017}
}

@article{jin2024non,
  title={Non-asymptotic global convergence analysis of BFGS with the Armijo-Wolfe line search},
  author={Jin, Qiujiang and Jiang, Ruichen and Mokhtari, Aryan},
  journal={Advances in Neural Information Processing Systems},
  volume={37},
  pages={16810--16851},
  year={2024}
}

@article{liu1989limited,
  title={On the limited memory BFGS method for large scale optimization},
  author={Liu, Dong C and Nocedal, Jorge},
  journal={Mathematical programming},
  volume={45},
  number={1},
  pages={503--528},
  year={1989},
  publisher={Springer}
}

@article{vu2025benchmarking,
  title={Benchmarking variants of the Adam optimizer for Quantum Machine Learning Applications},
  author={Vu, Tuan Hai and Le, Vu Trung Duong and Pham, Hoai Luan and Nakashima, Yasuhiko},
  journal={IEEE Open Journal of the Computer Society},
  year={2025},
  publisher={IEEE}
}

@article{hassan2023effect,
  title={The effect of choosing optimizer algorithms to improve computer vision tasks: a comparative study},
  author={Hassan, Esraa and Shams, Mahmoud Y and Hikal, Noha A and Elmougy, Samir},
  journal={Multimedia Tools and Applications},
  volume={82},
  number={11},
  pages={16591--16633},
  year={2023},
  publisher={Springer}
}

@article{semenov2025benchmarking,
  title={Benchmarking optimizers for large language model pretraining},
  author={Semenov, Andrei and Pagliardini, Matteo and Jaggi, Martin},
  journal={arXiv preprint arXiv:2509.01440},
  year={2025}
}

@article{kovacs2025mace,
  title={Mace-off: Short-range transferable machine learning force fields for organic molecules},
  author={Kov{\'a}cs, D{\'a}vid P{\'e}ter and Moore, J Harry and Browning, Nicholas J and Batatia, Ilyes and Horton, Joshua T and Pu, Yixuan and Kapil, Venkat and Witt, William C and Magdau, Ioan-Bogdan and Cole, Daniel J and others},
  journal={Journal of the American Chemical Society},
  volume={147},
  number={21},
  pages={17598--17611},
  year={2025},
  publisher={ACS Publications}
}

@book{nesterov2003introductory,
  author    = {Yurii Nesterov},
  title     = {Introductory Lectures on Convex Optimization: A Basic Course},
  series    = {Applied Optimization},
  volume    = {87},
  year      = {2003},
  publisher = {Springer},
  isbn      = {978-1-4419-8853-1}
}

@article{bottou2018optimization,
  author  = {L\'eon Bottou and Frank E. Curtis and Jorge Nocedal},
  title   = {Optimization Methods for Large‐Scale Machine Learning},
  journal = {SIAM Review},
  volume  = {60},
  number  = {2},
  pages   = {223--311},
  year    = {2018},
  doi     = {10.1137/16M1080173}
}

@article{amari1993backpropagation,
  title={Backpropagation and stochastic gradient descent method},
  author={Amari, Shun-ichi},
  journal={Neurocomputing},
  volume={5},
  number={4-5},
  pages={185--196},
  year={1993},
  publisher={Elsevier}
}

@article{soper2000radial,
  title={The radial distribution functions of water and ice from 220 to 673 K and at pressures up to 400 MPa},
  author={Soper, AK},
  journal={Chemical Physics},
  volume={258},
  number={2-3},
  pages={121--137},
  year={2000},
  publisher={Elsevier}
}

@article{hoy1979precise,
  title={A precise solution of the rotation bending Schr{\"o}dinger equation for a triatomic molecule with application to the water molecule},
  author={Hoy, AR and Bunker, Po R},
  journal={Journal of Molecular Spectroscopy},
  volume={74},
  number={1},
  pages={1--8},
  year={1979},
  publisher={Elsevier}
}

@article{ayappa2019enhancing,
  title={Enhancing the Dynamics of Water Confined between Graphene Oxide Surfaces with Janus Interfaces: A Molecular Dynamics Study.},
  author={Ayappa, K Ganapathy and others},
  journal={The journal of physical chemistry. B},
  volume={123},
  number={13},
  pages={2978--2993},
  year={2019}
}

@article{kresse1996efficient,
  title={Efficient iterative schemes for ab initio total-energy calculations using a plane-wave basis set},
  author={Kresse, Georg and Furthm{\"u}ller, J{\"u}rgen},
  journal={Physical review B},
  volume={54},
  number={16},
  pages={11169},
  year={1996},
  publisher={APS}
}

@article{gajdos2006linear,
  title={Linear optical properties in the projector-augmented wave methodology},
  author={Gajdo{\v{s}}, M and Hummer, Kerstin and Kresse, G and Furthm{\"u}ller, J and Bechstedt, FJPRB},
  journal={Physical Review B—Condensed Matter and Materials Physics},
  volume={73},
  number={4},
  pages={045112},
  year={2006},
  publisher={APS}
}

@article{togo2015first,
  title={First principles phonon calculations in materials science},
  author={Togo, Atsushi and Tanaka, Isao},
  journal={Scripta materialia},
  volume={108},
  pages={1--5},
  year={2015},
  publisher={Elsevier}
}

@article{mortazavi2020exploring,
  title={Exploring phononic properties of two-dimensional materials using machine learning interatomic potentials},
  author={Mortazavi, Bohayra and Novikov, Ivan S and Podryabinkin, Evgeny V and Roche, Stephan and Rabczuk, Timon and Shapeev, Alexander V and Zhuang, Xiaoying},
  journal={Applied Materials Today},
  volume={20},
  pages={100685},
  year={2020},
  publisher={Elsevier}
}

\end{document}